\documentclass[sn-mathphys,Numbered]{sn-jnl}

\usepackage{graphicx}%
\usepackage{multirow}%
\usepackage{amsmath,amssymb,amsfonts}%
\usepackage{amsthm}%
\usepackage{afterpage}
\usepackage{mathrsfs}%
\usepackage[title]{appendix}%
\usepackage{xcolor}%
\usepackage{textcomp}%
\usepackage{manyfoot}%
\usepackage{booktabs}%
\usepackage{algorithm}%
\usepackage{algorithmicx}%
\usepackage{algpseudocode}%
\usepackage{listings}%
\usepackage{flafter}%

\begin{document}

\title[Article Title]{Force Propagation in Active Cytoskeletal Networks}


\author*[1]{\fnm{Shichen} \sur{Liu}}\email{sliu7@caltech.edu}

\author[1]{\fnm{Rosalind Wenshan} \sur{Pan}}

\author[2]{\fnm{Heun Jin} \sur{Lee}}

\author[1]{\fnm{Shahriar} \sur{Shadkhoo}}

\author[1]{\fnm{Fan} \sur{Yang}}

\author[1]{\fnm{David} \sur{Larios}}

\author[3]{\fnm{Chunhe} \sur{Li}}

\author[3]{\fnm{Zijie} \sur{Qu}}

\author[1, 2]{\fnm{Rob} \sur{Phillips}}

\author*[1]{\fnm{Matt} \sur{Thomson}}\email{mthomson@caltech.edu}

\affil*[1]{\orgdiv{Division of Biology and Biological Engineering}, \orgname{California Institute of Technology}, \orgaddress{\city{Pasadena}, \postcode{91125}, \state{CA}, \country{USA}}}

\affil[2]{\orgdiv{Department of Applied Physics}, \orgname{California Institute of Technology}, \orgaddress{\city{Pasadena}, \postcode{91125}, \state{CA}, \country{USA}}}

\affil[3]{\orgdiv{UM-SJTU Joint Institute}, \orgname{Shanghai Jiao Tong University}, \orgaddress{\city{Shanghai}, \postcode{200240}, \state{Shanghai}, \country{China}}}


\abstract{In biological systems, energy-consuming active networks of motor and filament proteins generate the forces that drive processes, including motility, shape change, and replication. Despite the integral role of such active material, how molecular-scale interactions can enable the global organization and transmission of forces across micron length scales remains elusive. Here, we demonstrate that the bundling of microtubules can shift motor filament active matter between a global force-propagating phase and a local force-confined phase. An increase in the average microtubule bundle length from 0.49$\mu$m to 5.09$\mu$m results in a transition from a local to a global phase, characterized by an abrupt transition in correlation length from 57.4 $\mu$m to 247.5 $\mu$m. The global phase generates forces up to 26.1 pN that enable applications, including cell transport and droplet motility. Through theory and simulation, we demonstrate that even a minority species of long microtubules can induce a percolation transition between a local and global phase, providing a mechanism for the regulation of force transmission in cells. Our results reveal potential mechanisms underlying the propagation of forces in cells and enable the engineering of active materials in synthetic biology and soft robotics.}

\maketitle

\section*{Introduction}

In biological systems, nanometer-sized proteins like tubulin and actin self-assemble into filaments that interact with motor proteins to generate forces. This molecular-scale activity organizes into larger structures such as networks, asters, and bundles, spanning tens to hundreds of microns \cite{Huxley1954-mu, Mitchison1984-rl, Shelley2016-qf}. These multi-scale organizations enable critical cellular processes, from intracellular transport to cell division. Molecular-scale forces transmit and amplify through space as proteins consume energy and self-organize into structures up to a million-fold bigger than a single molecule \cite{Howard2001-kf, Shelley2016-qf, Heisenberg2013-qr, Oakes2014-dm}. Exactly how the molecular interactions enable force propagation from nanometer to micrometer length scales, however, remains elusive. The mitotic spindle, the cytoskeletal structure that pulls apart chromatids during mitosis, spans from a few microns in yeast to 50 microns in \textit{Xenopus} oocytes and is capable of exerting force pulling chromatids over a few to ten microns. The lamellipodium, a cytoskeletal projection on the leading edge of a cell, spans over tens of micrometers and propagates forces across the leading edge to enable cell motility \cite{Heald1997-gp, Biswas2021-ph, Desai1998-xx, Gard1992-qa, Heisenberg2013-qr, Abercrombie1970-id, Li2023-ev}. Force propagation is also apparent in \textit{Drosophila melanogaster} oocytes, where studies of streaming during oocyte development have revealed a transition from a spatially disordered cytoskeleton, supporting only micron-scale movements, to an ordered state with cell-spanning vortical flow \cite{Theurkauf1992-ml, Goldstein2015-tk, He2011-re, Bastock2008-mo, Stein2021-wv}. The knowledge gap between how self-organizing filaments and motor proteins influence force propagation and force dissipation limits our ability to harness these mechanisms in engineering biological and synthetic systems.

Reconstituted active cytoskeletal networks of purified motor and filament proteins enable molecular-level analysis of force propagation in cytoskeletal systems \cite{Sanchez2012-jp, Nedelec1997-rw, Ross2019-ol}. These systems allow us to observe how microtubule polymerization and bundling kinetics evolve over time, dramatically altering network properties. Microtubule polymerization dynamics, characterized by growth and catastrophe rates, directly influence the length distribution of filaments \cite{Mitchison1984-rl}. This distribution, in turn, affects the network's mechanical properties \cite{Gardel2008-xz}. Bundling, facilitated by motor proteins or crosslinkers, increases the effective rigidity of microtubule structures and can extend force transmission distances by an order of magnitude \cite{Needleman2004-fu, Portran2013-da, Baas2016-kf, Tolic-Norrelykke2008-jg, Kapitein2015-sm, Forth2017-mv}. While cellular organization involves numerous interacting components, these simplified systems facilitate identification of key biophysical principles governing force transmission over ten to hundred micron scales \cite{Yang2022-uh}. Elucidating how these time-dependent processes determine long-range force transmission is crucial for understanding cellular processes such as spindle formation, intracellular transport, and tissue morphogenesis \cite{Brugues2014-kc, Sanchez2005-tb, Brangwynne2006-my, Mitchison1984-rl, Whitesides2015-ca}.
 
In this study, we use light-controlled active matter to probe microscopic interactions that determine the length scale of force propagation in motor filament active matter.\cite{Ross2019-ol} We demonstrate that the microtubule bundle length distribution controls a transition between a local force-confined phase with correlation lengths below 57.4 $\mu$m and a global force-propagation phase with correlation lengths above 247.5 $\mu$m. Through simulation and theory, we show that a small population of long microtubules is sufficient to induce long-range force transmission. By controlling the transition between force dissipation and propagation, we enable the modulation of active-matter-powered material transport and manipulation. This fine-tuned control over force transmission opens the door for light-controlled active matter-based cell sorting applications and innovative approaches to study and manipulate cellular events.

\section*{Main} 
\subsection*{Active matter incubation time switches contraction between the global force-propagating phase and the local force-confined phase.}

During the cycles of growth, reconfiguration, and division, cells switch their internal cytoskeletal networks between phases of long-range and short-range organization through modulation of the length and bundling effect of microtubules \cite{Kapitein2015-sm, Siegrist2007-xl, Dogterom2013-ko, Petry2013-il, Brugues2012-qi}. To investigate the transition between force-confined and force-propagating properties of the two cytoskeletal organizations, we used a defined and optically controlled in vitro experimental system composed of stabilized microtubules and engineered kinesin motor proteins \cite{Nedelec1997-rw, Kruse2004-vq, Sanchez2012-jp, Hentrich2010-vp, Doostmohammadi2018-td}. Our microtubule kinesin active matter system consists of stabilized microtubules labeled with fluorophore and kinesin motor proteins fused with optically dimerizable iLID (improved light-induced dimer) proteins \cite{Ross2019-ol}. Upon light activation, the kinesin protein dimerization induces interactions between kinesins and neighboring microtubules, which leads to microtubule network formation and contraction into asters. In previous work, we observed that pre-incubation of motors and microtubules could induce a transition between long-range force-propagating and short-range force-confined phases. This transition likely stems from changes in microtubule bundling, a key factor in determining network properties and force transmission \cite{Sanchez2012-jp, Hentrich2010-vp}. Bundles can significantly alter the effective length and rigidity of microtubule structures \cite{Needleman2004-fu}, potentially serving as a mechanism for switching between organizational phases. Our in-vitro system allows us to isolate and study how bundle formation influences force propagation in a controlled environment. Therefore, we performed a systematic study of system organization while quantitatively varying the duration of motor-microtubule incubation time. 

We investigated how pre-activation incubation time affects cytoskeletal network organization in a light-controlled active matter system. Our experiments revealed two distinct phases of microtubule organization and dynamics. For incubation times below 40 minutes, we observed a local phase characterized by small microtubule asters smaller than 50 $\mu$m confined within the light-activated region. In contrast, incubation times exceeding 65 minutes led to a global phase with long-range effects, forming structures up to 800 $\mu$m and recruiting microtubules from over 300 $\mu$m away from the light activated region \cite{Yang2022-uh}. We established a calibration curve by calculating the average intensities at different microtubule densities based on tubulin concentration in our system. Quantitative analysis showed that the global phase outperforms the local phase in material concentration and transport. The global phase exhibited an increase in microtubule density from 1 MT/$\mu$m$^2$ to 3.4 MT/$\mu$m$^2$ within the light-activated region, compared to a small increase from 1 MT/$\mu$m$^2$ to 1.3 MT/$\mu$m$^2$ in the local phase (Figure 1B and Supplemental Figure 3). Moreover, the global phase induced large-scale microtubule buckling and long-range material recruitment, capabilities absent in the local phase. These findings demonstrate that extended incubation enables the emergence of coordinated, system-wide behaviors in active microtubule networks, significantly enhancing their ability to concentrate and organize material over large distances.

To determine how far these movements to concentrate materials propagate through the system, we quantified the coherent motion length scales. We calculated the Eulerian two-point correlation coefficient (Figure 1C) and the corresponding correlation length (Figure 1D) \cite{Chen2019-ke}. This spatial correlation coefficient reveals the persistence of velocity vector directions and magnitudes across space. We computed the correlation coefficient by anchoring the first point at the rightmost edge of the rectangular light pattern and varying the second point's distance horizontally along the light-activated region. For the local phase, we fitted $A \cdot \exp(-x / l)$ to the correlation coefficient data, where $l$ represents the correlation length. For the global phase, due to its complex behavior and negative correlation, we defined the correlation length as the characteristic length scale at which the correlation coefficient decayed by $1/e$.

Our analysis revealed two distinct correlation lengths for the global and local phases. The local phase, observed in experiments with incubation times under 40 minutes, exhibited correlation lengths clustering below 57.4 $\mu$m. This aligns with our observations of small asters with uncorrelated motions. We observed an abrupt increase in correlation length from 57.4 $\mu$m to 247.5 $\mu$m at 65-minute incubation time, marking the shift from local to global phase. Within the global phase, correlation lengths increased gradually from 247.5 $\mu$m to 411.3 $\mu$m with longer incubation times. Hence, we classify the local phase as having correlation length below 80 $\mu$m and the global phase as having correlation length above 230 $\mu$m. We did not observe any further discontinuous jumps once the system entered the global phase. The local phase exhibits a sharp decline in the correlation coefficient as inter-point distance increases, eventually fluctuating around zero at 60 $\mu$m. In the global phase, we observed persistently correlated motion over long distances, with the correlation coefficient becoming negative at 300-500 $\mu$m, approximately half the length of the microtubule network. This transition to negative correlation indicates opposing velocity vectors, reflecting symmetric material recruitment. 

To quantify the global phase's capacity for persistent material recruitment into the light-activated region, we defined an elliptical region of interest (ROI) around the geometric center. This ROI allowed us to measure microtubule transport over time [Figure 1E]. We overlaid the cumulative mass flux along the ROI boundary onto the final frame of a 10-minute (local phase) and a 170-minute (global phase) incubation experiment, along with the velocity field derived from optical flow analysis [Supplemental Figure 1]. Consistent with our observations, the local phase recruited significantly less material compared to the global phase. The majority of mass flux occurred through the left and right vertices of the ROI. We also calculated the total mass flux along the ROI boundary over time [Figure 1F]. Initially (0-3 minutes), all experiments in the global phase and some experiments in the local phase exhibited similar magnitudes of material transport into the ROI, attributed to local and global aster contraction and movement at the ROI vertices. As experiments progressed, the global phase demonstrated persistent microtubule transport into the ROI, while the local phase mass flux remained near zero.
\begin{figure}[htbp]
    \centering
    \includegraphics[width=\textwidth]{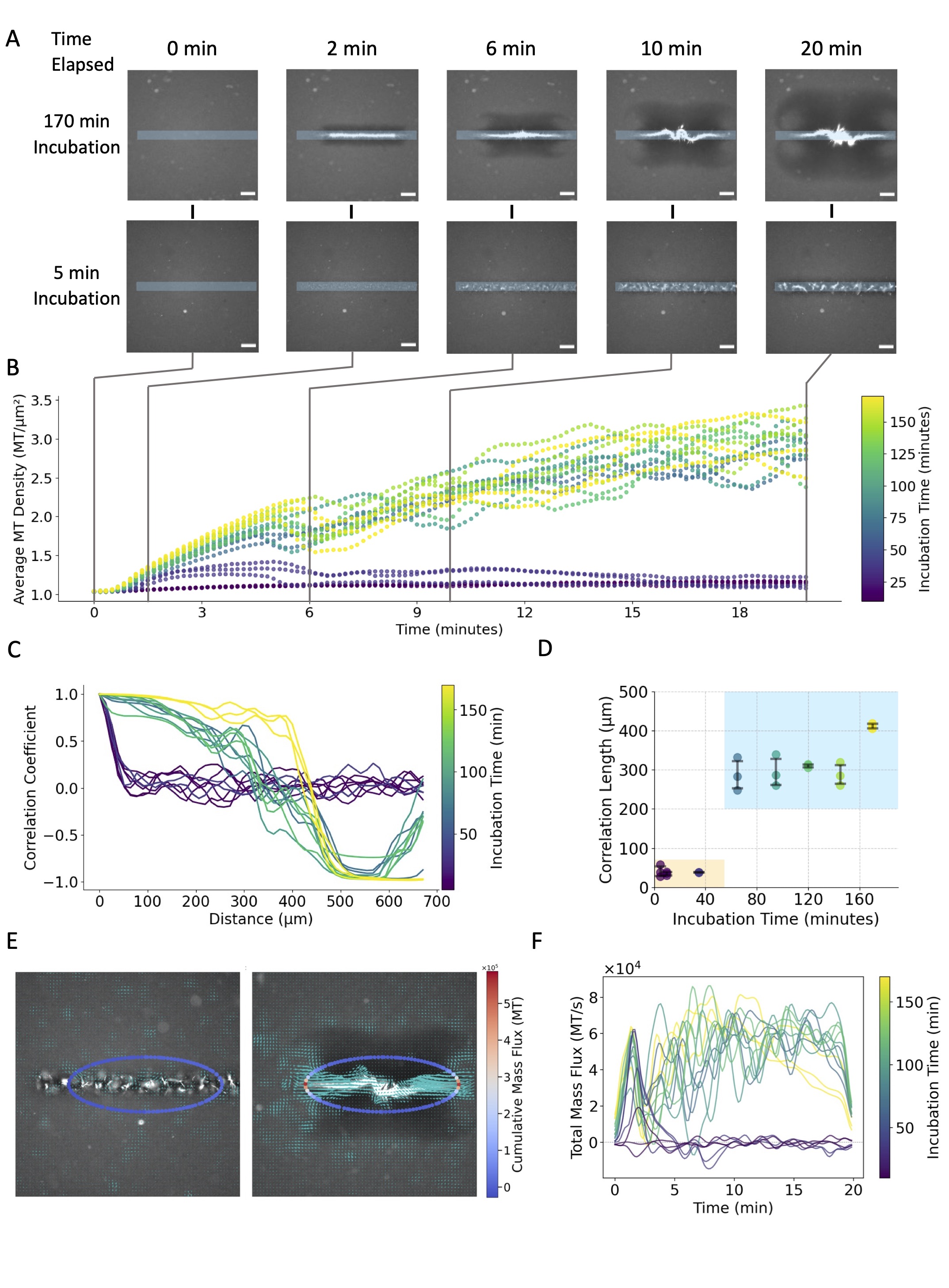}
    \label{fig:your-figure}
\end{figure}

\begin{figure}[htbp]%
\centering
\caption{\textbf{Microtubule Incubation-Induced Phase Transition Between Local and Global Phases
}. \textbf{A}, Images of labeled microtubules during aster assembly. The top row shows the global contraction phase with an incubation time of 170 minutes, while the bottom row shows the local contraction phase with an incubation time of 5 minutes. Scale bar: 100 $\mu$m. \textbf{B}, Microtubule density in the light-illuminated region, color-coded by incubation time. We observed increases in microtubule density in the light illuminated region as time progresses in the experiments with higher incubation time, while microtubule density stayed at 1-1.3 for experiments with lower incubation time.
  \textbf{C}, Euclidean two-point correlation coefficient of the light-illuminated region, color-coded by incubation time. We computed the correlation coefficient by fixing a point at the right most point of the light illuminated region and varied the second point at different distances. Experiments with incubation time over 65 minutes showed persistent correlated movement within the light activated region, and a negative correlation representing directional change of velocity changes. While the experiments with incubation time below 65 minutes showed a sharp drop-off at 50 $\mu$m showing no persistent correlated movements.
\textbf{D}, Correlation length computed from the correlation coefficient function. The orange shaded region represents the local contraction phase, and the blue region represents the global contraction phase. Correlation length $l$ of the local phase is calculated by fitting $A \cdot \exp(-x / l)$ to the correlation coefficients. Correlation length $l$ of the global phase is calculated by finding the distance between two points when the correlation coefficients decayed by $1/e$
\textbf{E}, Mass flux in the region of interest overlaid on representative velocity fields of the local (left) and global (right) phases of contraction. For each boundary point $i$ at time $t$, we calculate the flux $F_i(t)$ as: $F_i(t) = \pm v_x(i,t) \cdot \rho(i,t) \cdot \Delta x$
\textbf{F}, Total mass flux at the region of interest, with data color-coded by incubation time.}
\label{fig1}
\end{figure}

\subsection*{Increases in microtubule length induce a percolation transition.} To elucidate the microscopic mechanism underlying the transition from the local force-confined phase to the global force-propagating phase, we performed fluorescence recovery after photobleaching (FRAP) experiments. FRAP allows for the measurement of the effective length of microtubule structures from the diffusion coefficient of the recovery of fluorescent microtubules after photobleaching a specific region. We found that the length of the microtubule structures increased linearly from 0.9 $\mu$m to 5 $\mu$m over a period of 2 hours, with a rate of 2.69 $\mu$m/hr at room temperature(Figure 2A). This rate of increase is 4 times higher than the GMP-cpp (Supplemental Information) polymerization rate of tubulin, which is 0.5 $\mu$m/hr at room temperature \cite{Najma2023-ya}. GMP-cpp is a slowly hydrolyzable analog of GTP that stabilizes microtubules, and its polymerization rate serves as a reference for the observed growth rate. Thus, the higher growth rate observed in our experiments suggests that microtubules exhibit bundling behavior, which increases the length of each microtubule structure by 5-fold in 120 minutes.

To test the material transport capability of the contractions, we placed 10 $\mu$m polystyrene tracer beads in the system and measured their movements. The tracer beads underwent Brownian motion for incubation times of 0 to 30 minutes, indicating that the active matter system could not exert significant forces on the cargo. However, after incubating the system for more than 60 minutes, the contraction moved the tracer beads by up to 200 $\mu$m, and the cumulative force exerted on the beads increased by up to 25-fold to 26 pN (Figure 2b). The cumulative force was calculated by analyzing the displacement of the beads and using Stokes' law to estimate the force required to overcome the drag in the medium. The directed movement of the tracer beads demonstrates that bundled microtubules drive the global force-propagating phase, exhibiting material transport and long-range order.

To validate the microtubule-bundling-induced transition between the local force-confined phase and the global force-propagating phase, we first employed Cytosim, a numerical simulation platform designed to model cytoskeletal mechanics by simulation each microtubule and kinesin (Figure 2c) \cite{Nedelec2007-ln}. We investigated whether increasing the length of the microtubules would cause the system to transition from having many local asters to having a global aster. The length of the microtubules was modeled using a modified gamma distribution, where the shortest microtubule was 1 $\mu$m, and the longest was 25 $\mu$m \cite{Odde1995-gb}. We fixed the total length of microtubules in simulations to 1000 $\mu$m, so as microtubules grow longer, the total number of microtubules decreases. As we increased the scale parameter, which introduced a small portion of long microtubules, the simulated active matter system transitioned from having many local asters to having a global aster. When the scale parameter of the microtubule length distribution increased to 2.75, with an average microtubule length of 3.7 $\mu$m, we observed plateaus in both the fraction of microtubules in the largest aster and the mean distance between any pair of microtubules, indicating the system's transition into the global phase (Figure 2e). Increasing either or both the shape parameter, which primarily influences the peak of the length distribution, and the scale parameter, which primarily influences the spread of the length distribution, caused the simulated system to transition from the local phase with many local asters to the global phase where most of the microtubules are in one aster (Figure 2g). Our numerical simulation shows that as the microtubule length increases, the active matter system transitions from the local to the global phase.

To further validate the phase transition, we developed a more generalized network-based simulation rooted in percolation theory (Figure 2d). Percolation theory describes the formation of connected clusters in a random graph and has been applied to study phase transitions in various systems \cite{Artime2021-iv}. In our network-based simulation, instead of explicitly representing each microtubule and kinesin, we abstracted the system into nodes, representing the center of mass of microtubules, and edges, symbolizing potential connections between them. The edge probabilities, reflecting the likelihood of interaction based on the relative lengths and spatial orientations of the microtubules, were determined using gamma distribution. We first explored varying the scale parameter while keeping the shape parameter at 1. As we increased the scale parameter, we observed an exponential increase in the size of the largest connected component, followed by a plateau where all the nodes became connected to each other (Figure 2f). When increasing both the scale and shape parameters, the simulated system exhibited a transition from the disconnected phase to the connected phase with a giant component, resembling a percolation transition (Figure 2h). This phase transition in the network-based simulation validates the microtubule-bundling-mediated local force-confined to the global force-propagating phase transition observed in the experimental system. Furthermore, our simulations hint at the influential role of MT length distribution in facilitating spatial dynamics within the cytoskeleton. This aligns with the observed load-bearing attributes of microtubule networks, attributed to their lateral interactions with other filaments and molecular structures \cite{Brangwynne2006-my}.
\begin{figure}[htbp]
\centering
\includegraphics[width=0.9\textwidth]{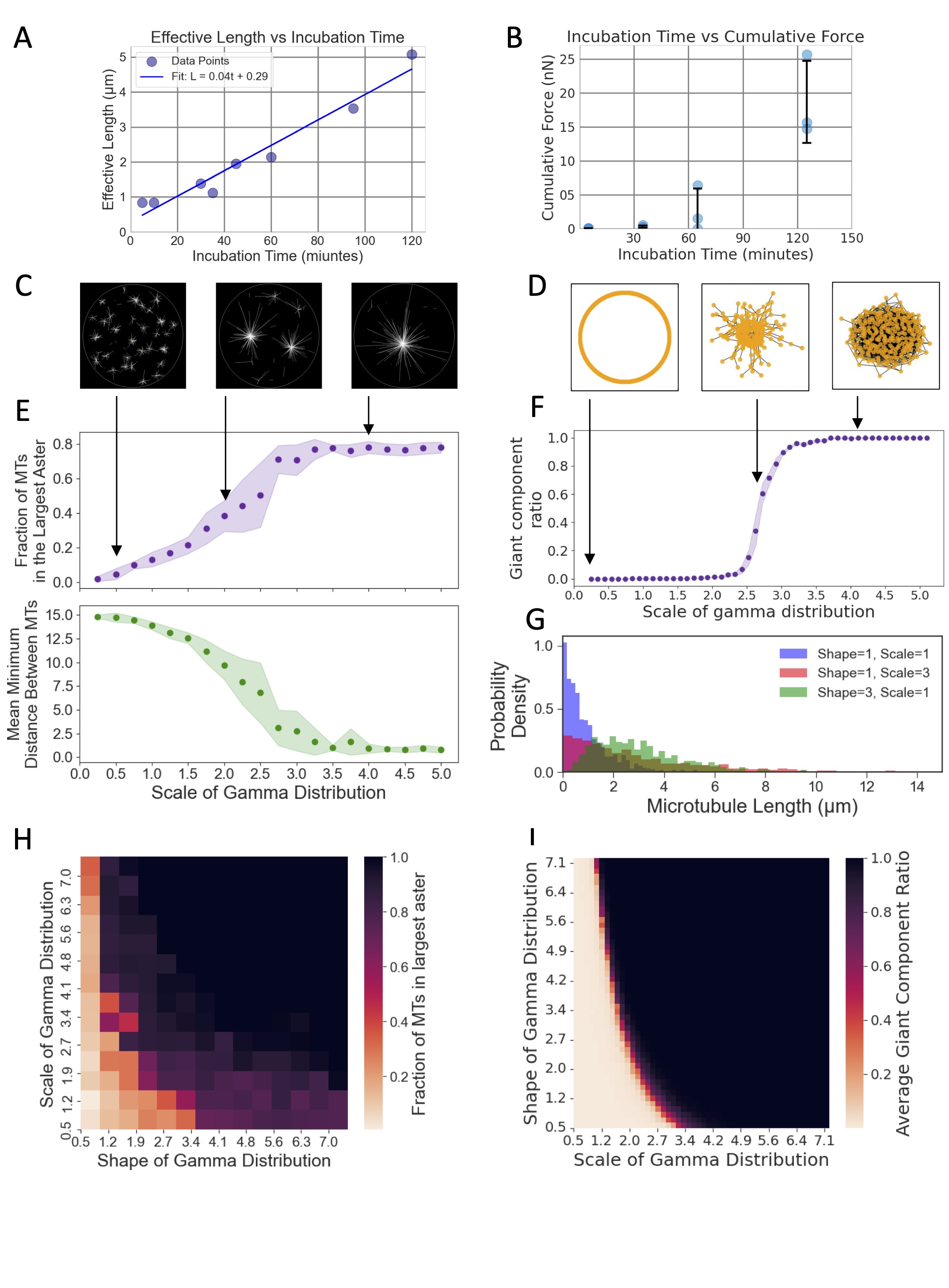}
\end{figure}

\begin{figure}[htbp]

\caption{\textbf{A fivefold increase in microtubule bundle length enables force propagation phase in active networks.}. \textbf{A}, Incubation time increases microtubule effective length at the rate of 2.69 $/mu$m per hour measured by Fluorescence Recovery After Photobleaching (FRAP) \textbf{B}, Active matter system exert close to 100 times more force after incubation by moving polystyrene tracer beads. Forces calculated by $F = 6 \pi \eta r v$, where $v$ is measured using particle tracking. N = 15 beads for 5 minutes incubation, N = 17 beads for 35 minutes incubation, N = 18 beads for 65 minutes incubation, N = 14 beads for 125 minutes incubation, n = 3 independent experiments for each incubation time\textbf{C}, Images of representative Cytosim simulations under different microtubule length distribution. \textbf{D}, Images of representative percolation simulation under different microtubule length distribution.  \textbf{E},  The top plot shows the fraction of microtubules in the largest aster versus the scale parameter of the gamma distribution. The bottom plot shows the mean of the minimum distance between each pair of microtubules versus the scale parameter of the gamma distribution. Shape = 1 \textbf{F}, Giant component ratio vs scale of gamma distribution for the percolation simulations. Shape = 1 \textbf{G}, Representative gamma distribution for simulated microtubule length. \textbf{H}, Heatmap showing the relationship of the shape and scale of gamma distributed microtubule species and fraction of microtubules in the largest aster. \textbf{I}, Heatmap showing the relationship of the shape and scale of gamma distributed edge species and the giant component ratio.}
\label{fig2}
\end{figure}

\subsection*{The global force propagating phase enables persistent material transport} 
To study the long-range interactions enabled by the global force-propagating phase of the active cytoskeleton network, we constructed an aster merger operation (Figure 3a) where we connected two asters using light patterns, as previously described by \cite{Ross2019-ol}. In both the local and global phases, two asters were formed by circular light-excitation patterns, with the local phase aster being smaller and depleting less material compared to the global phase aster. Upon connecting the pair of asters using a rectangular pattern with a high aspect ratio, we observed that in the global phase, the two asters merged after a short delay time (~ 1 minute), while the aster pair in the local phase did not exhibit any movement. These findings further reinforced that the moving and merging global phase asters are dynamic and constantly remodeling, whereas the local phase asters are closer to steady-phase structures \cite{Surrey2001-nl}.

The ability to direct aster formation and movement, both spatially and over time, marks a critical progression toward transport applications using the active cytoskeleton network. We demonstrated the simultaneous movement of multiple asters using dynamic light patterns (Figure 3b). As asters moved, inflows of microtubule bundles emerged within the light pattern, feeding into and pulling the aster behind. Concurrently, outflows, resembling comet-tail streams, trailed the moving asters. During the aster movement, the global phase asters followed the dynamic light patterns without interruption. In contrast, local phase asters struggled to match the pace of the light patterns, often leading to transient and static asters that subsequently gave rise to new aster formations along the light pattern's path.

\begin{figure}[htbp]%
\centering
\includegraphics[width=0.9\textwidth]{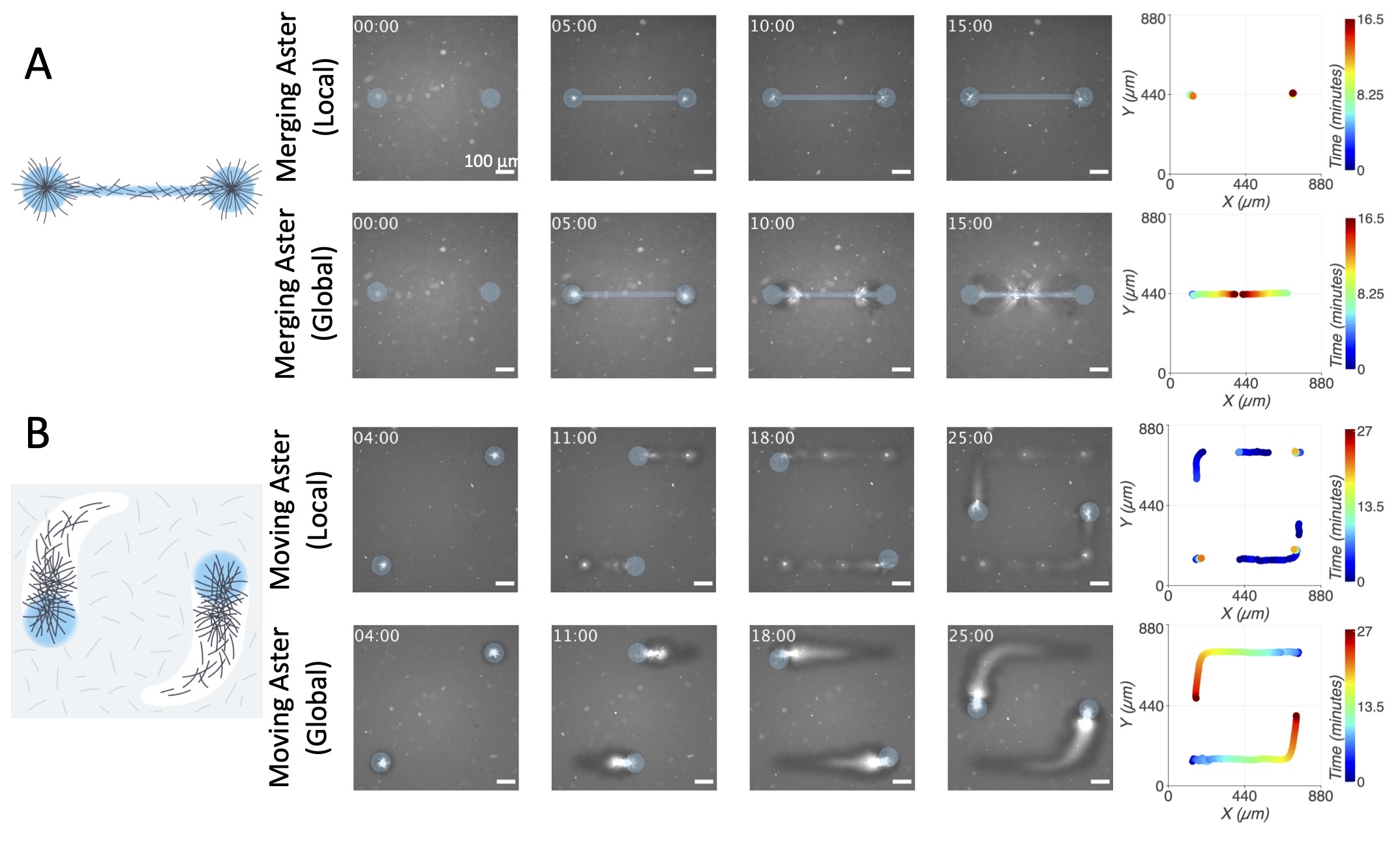}
\caption{\textbf{Force propagation driven by global contraction enables transport of microtubules}. \textbf{A}, Aster linking in both local (top) and global (bottom) phases. Plots on the right shows trajectories of asters position using image thresholding. The asters in the local phase do not exhibit movement when linking them with an additional thin light pattern, while the two asters in the global phase merged. \textbf{B},  Aster moving in both local (top) and global (bottom) phases. The plots on the right shows the time each aster persists. Asters in the local phase did not follow the light pattern, thus many new asters emerged as the dynamic light pattern moved. Asters in the global phase followed the light pattern throughout the entire movement of the dynamic light pattern. Time unit: minutes}
\label{fig4}
\end{figure}

\subsection*{Force propagating active matter powers cell transporter and droplet motility}

The active network's ability to generate force and transport materials suggests its potential as an engine and transport agent for biological processes and future bio-robots. To demonstrate this capability with biological materials, we investigated the movement of suspended human Jurkat cells within the active cytoskeleton network.Upon light activation, a contracting global phase aster formed and began capturing cells. Using dynamic light patterns (Figure 4A), we directed the aster to move 1 mm while collecting cells along its path (Figure 4B). As the aster moved, its cross-sectional area increased linearly (Figure 4C), aiding in cell retention. When approaching a cell, the aster's inflows generated power in the range of 10 to 59 nW, driving the cell toward its center (Figure 4D). The global phase aster demonstrated the ability to capture multiple cells simultaneously. Initially, it only captured cells within the light-activated region. However, as incubation time and aster size increased, material outside this region was rapidly recruited, expanding the aster's capture range. Of 32 cells captured, only 2 were released during transport, both from the aster's "arms." Cells in the aster's core were consistently transported throughout the experiment, suggesting structural differences between the "arms" and "core." The core's randomly crosslinked structure may provide greater carrying capacity compared to the arms. Notably, uncaptured cells showed minimal movement compared to captured cells (Figure 4E) The precise control and localized force generation capabilities of our active matter system, operating in the 10-59 nW range, making it well-suited for delicate tasks such as single-cell manipulation, enabling the gentle transport or positioning of individual cells within microfluidic devices without the risk of photodamage or electrical disruption from other methods \cite{Xie2020-jc, Eriksson2007-og}.

In the biological context, the cytoskeleton and the active fluids generated by it have been proposed to lead to the emergence of cytoplasmic streaming, which can enhance cellular transport. To mimic this biological process, we encapsulated our active cytoskeleton network in aqueous droplets emulsified in oil. When activated by light, the active cytoskeleton network did not seem to propel the aqueous droplet in either the global phase or the local phase (Figure 4f). However, outside of our current defined phases, we were able to generate directed movement of the droplet by light activation, and the movement quickly stopped when we turned off the light. In this phase, the active network does not contract into global or local asters but exhibits a highly crosslinked state with minimal contraction. We hypothesize that the crosslinked network created localized tension at the water-oil interface, which resulted in Marangoni flow and propelled the droplet forward.
\begin{figure}[H]%
\centering
\includegraphics[width=0.9\textwidth]{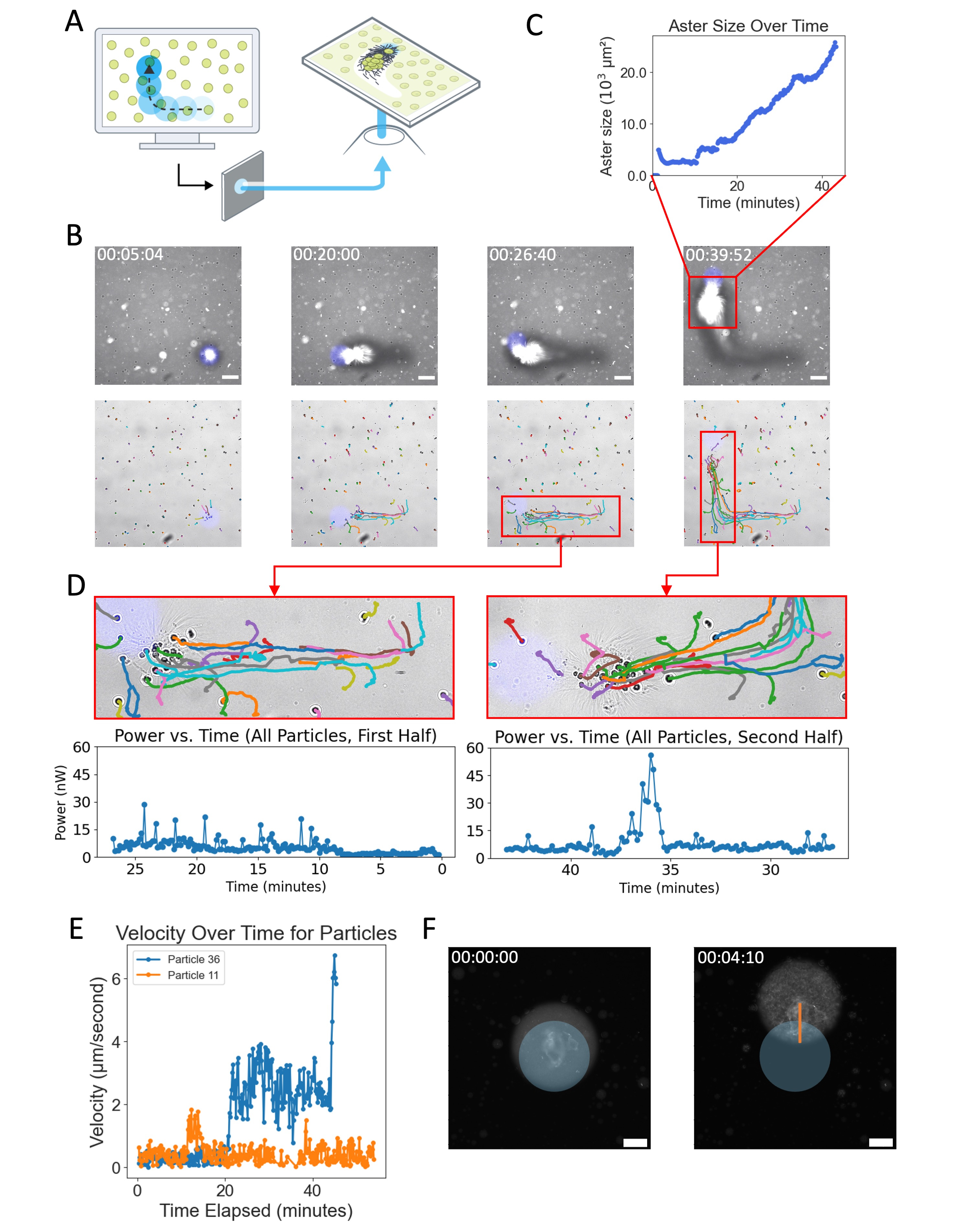}
\caption{\textbf{Aster-based active cell transport}. \textbf{A}, Schematics of cell transport use active matter. Briefly, in a field with active matter, and cells, we utilized a dynamic light pattern similar to the one in Figure 3B to recruit and transport the cells using global phase asters \textbf{B},  Time-lapsed images of using aster (top) to concentrate and transport Jurkat T cells (bottom). Scale bar, 100$\mu$m.\textbf{C}, Aster area increases as aster displacement increases due to constant recruitment of microtubules. \textbf{D}, Power generated by moving cells with respect to time. Briefly, we calculated the force using the same method as in Figure 2B by using the Stokes equation and cell tracking. Power is calculated by $P = W/\Delta t$ \textbf{E}, Example of cell velocity of a stationary cell (orange) vs a cell captured by the aster (blue). \textbf{F}, Proof of concept motile droplet powered by contracting active cytoskeleton. Blue light indicates the light illuminated region and orange line represents the trajectory of the droplet. Scale bar, 100 $\mu$m}
\label{fig5}
\end{figure}

\section*{Summary and Outlook}
Our study quantifies how time-dependent changes in microtubule length distribution drive a phase transition in active cytoskeletal networks, shifting from a local force-confined phase to a global force-propagating phase. By varying incubation time, we observe a linear increase in the effective length of microtubule structures at a rate of 2.05 $\mu$m/hr, resulting in a 5-fold increase from 0.9 $\mu$m to 5 $\mu$m over 120 minutes. This increase in length correlates with significant changes in the collective movement and force transmission properties of the active network. We observe a 3-fold jump in correlation length, from 80 $\mu$m to 240 $\mu$m, and a 20-fold increase in force magnitude, reaching up to 26 pN. The transition to the global force-propagating phase occurs at approximately 65 minutes of incubation, marked by an abrupt increase in correlation length from 80 $\mu$m to 240 $\mu$m.

While both polymerization and bundling could contribute to this length increase, we hypothesize that bundling plays a dominant role. The observed growth rate of 2.69 $\mu$m/hr at room temperature significantly exceeds the standard GMP-cpp polymerization rate of 0.5 $\mu$m/hr at the same tubulin and GMP-cpp concentration, suggesting that additional mechanisms beyond simple polymerization are at play. However, it is important to note that our experimental setup does not allow direct observation of individual microtubules, limiting our ability to definitively characterize the molecular-scale changes underlying this phenomenon.

To validate our experimental observations and explore the underlying mechanisms, we employed both Cytosim and percolation-based simulations. Our Cytosim model demonstrates that increasing the scale parameter of the microtubule length distribution to 2.75, corresponding to an average length of 3.7 $\mu$m, triggers a transition to a globally connected state. This aligns with our experimental findings and is further corroborated by our percolation simulations. The percolation model shows a rapid increase in the size of the largest connected component as the scale parameter increases, mirroring the phase transition observed experimentally. Importantly, both simulation approaches reveal that even a small population of long microtubules is sufficient to induce long-range force transmission. 

In the global phase (greater than 65 minutes incubation), we demonstrated potential active-matter-based cell transport method. The linear increase in aster cross-sectional area during movement suggests a time-dependent, scalable mechanism for material capture and transport, likely enhanced by induced fluid flows. These findings provide a quantitative link between molecular-scale structural changes and cellular-scale force propagation, illustrating how microtubule polymerization and bundling can lead to large-scale transitions in network behavior. We have successfully demonstrated the ability of our active cytoskeleton network to transport cells over millimeter-scale distances, induce physical stress on cells, and move droplets in a directed fashion. This capability opens tangible avenues for advancements in biological and medical applications \cite{Yang2022-uh}, potentially enabling new approaches in tissue engineering, drug delivery, and biomaterials design.

\bibliography{reference}


\begin{thebibliography}{47}
\ifx \bisbn   \undefined \def \bisbn  #1{ISBN #1}\fi
\ifx \binits  \undefined \def \binits#1{#1}\fi
\ifx \bauthor  \undefined \def \bauthor#1{#1}\fi
\ifx \batitle  \undefined \def \batitle#1{#1}\fi
\ifx \bjtitle  \undefined \def \bjtitle#1{#1}\fi
\ifx \bvolume  \undefined \def \bvolume#1{\textbf{#1}}\fi
\ifx \byear  \undefined \def \byear#1{#1}\fi
\ifx \bissue  \undefined \def \bissue#1{#1}\fi
\ifx \bfpage  \undefined \def \bfpage#1{#1}\fi
\ifx \blpage  \undefined \def \blpage #1{#1}\fi
\ifx \burl  \undefined \def \burl#1{\textsf{#1}}\fi
\ifx \doiurl  \undefined \def \doiurl#1{\url{https://doi.org/#1}}\fi
\ifx \betal  \undefined \def \betal{\textit{et al.}}\fi
\ifx \binstitute  \undefined \def \binstitute#1{#1}\fi
\ifx \binstitutionaled  \undefined \def \binstitutionaled#1{#1}\fi
\ifx \bctitle  \undefined \def \bctitle#1{#1}\fi
\ifx \beditor  \undefined \def \beditor#1{#1}\fi
\ifx \bpublisher  \undefined \def \bpublisher#1{#1}\fi
\ifx \bbtitle  \undefined \def \bbtitle#1{#1}\fi
\ifx \bedition  \undefined \def \bedition#1{#1}\fi
\ifx \bseriesno  \undefined \def \bseriesno#1{#1}\fi
\ifx \blocation  \undefined \def \blocation#1{#1}\fi
\ifx \bsertitle  \undefined \def \bsertitle#1{#1}\fi
\ifx \bsnm \undefined \def \bsnm#1{#1}\fi
\ifx \bsuffix \undefined \def \bsuffix#1{#1}\fi
\ifx \bparticle \undefined \def \bparticle#1{#1}\fi
\ifx \barticle \undefined \def \barticle#1{#1}\fi
\bibcommenthead
\ifx \bconfdate \undefined \def \bconfdate #1{#1}\fi
\ifx \botherref \undefined \def \botherref #1{#1}\fi
\ifx \url \undefined \def \url#1{\textsf{#1}}\fi
\ifx \bchapter \undefined \def \bchapter#1{#1}\fi
\ifx \bbook \undefined \def \bbook#1{#1}\fi
\ifx \bcomment \undefined \def \bcomment#1{#1}\fi
\ifx \oauthor \undefined \def \oauthor#1{#1}\fi
\ifx \citeauthoryear \undefined \def \citeauthoryear#1{#1}\fi
\ifx \endbibitem  \undefined \def \endbibitem {}\fi
\ifx \bconflocation  \undefined \def \bconflocation#1{#1}\fi
\ifx \arxivurl  \undefined \def \arxivurl#1{\textsf{#1}}\fi
\csname PreBibitemsHook\endcsname

\bibitem[\protect\citeauthoryear{Huxley and Hanson}{1954}]{Huxley1954-mu}
\begin{barticle}
\bauthor{\bsnm{Huxley}, \binits{H.}},
\bauthor{\bsnm{Hanson}, \binits{J.}}:
\batitle{Changes in the cross-striations of muscle during contraction and stretch and their structural interpretation}.
\bjtitle{Nature}
\bvolume{173}(\bissue{4412}),
\bfpage{973}--\blpage{976}
(\byear{1954})
\end{barticle}
\endbibitem

\bibitem[\protect\citeauthoryear{Mitchison and Kirschner}{1984}]{Mitchison1984-rl}
\begin{barticle}
\bauthor{\bsnm{Mitchison}, \binits{T.}},
\bauthor{\bsnm{Kirschner}, \binits{M.}}:
\batitle{Dynamic instability of microtubule growth}.
\bjtitle{Nature}
\bvolume{312}(\bissue{5991}),
\bfpage{237}--\blpage{242}
(\byear{1984})
\end{barticle}
\endbibitem

\bibitem[\protect\citeauthoryear{Shelley}{2016}]{Shelley2016-qf}
\begin{barticle}
\bauthor{\bsnm{Shelley}, \binits{M.J.}}:
\batitle{The dynamics of microtubule/motor-protein assemblies in biology and physics}.
\bjtitle{Annu. Rev. Fluid Mech.}
\bvolume{48},
\bfpage{487}--\blpage{506}
(\byear{2016})
\end{barticle}
\endbibitem

\bibitem[\protect\citeauthoryear{Howard}{2001}]{Howard2001-kf}
\begin{bbook}
\bauthor{\bsnm{Howard}, \binits{J.}}:
\bbtitle{Mechanics of Motor Proteins and the Cytoskeleton},
\bedition{New edition} edn.
\bpublisher{Sinauer Associates is an imprint of Oxford University Press}, \blocation{???}
(\byear{2001})
\end{bbook}
\endbibitem

\bibitem[\protect\citeauthoryear{Heisenberg and Bella{\"\i}che}{2013}]{Heisenberg2013-qr}
\begin{barticle}
\bauthor{\bsnm{Heisenberg}, \binits{C.-P.}},
\bauthor{\bsnm{Bella{\"\i}che}, \binits{Y.}}:
\batitle{Forces in tissue morphogenesis and patterning}.
\bjtitle{Cell}
\bvolume{153}(\bissue{5}),
\bfpage{948}--\blpage{962}
(\byear{2013})
\end{barticle}
\endbibitem

\bibitem[\protect\citeauthoryear{Oakes et~al.}{2014}]{Oakes2014-dm}
\begin{barticle}
\bauthor{\bsnm{Oakes}, \binits{P.W.}},
\bauthor{\bsnm{Banerjee}, \binits{S.}},
\bauthor{\bsnm{Marchetti}, \binits{M.C.}},
\bauthor{\bsnm{Gardel}, \binits{M.L.}}:
\batitle{Geometry regulates traction stresses in adherent cells}.
\bjtitle{Biophys. J.}
\bvolume{107}(\bissue{4}),
\bfpage{825}--\blpage{833}
(\byear{2014})
\end{barticle}
\endbibitem

\bibitem[\protect\citeauthoryear{Heald et~al.}{1997}]{Heald1997-gp}
\begin{barticle}
\bauthor{\bsnm{Heald}, \binits{R.}},
\bauthor{\bsnm{Tournebize}, \binits{R.}},
\bauthor{\bsnm{Habermann}, \binits{A.}},
\bauthor{\bsnm{Karsenti}, \binits{E.}},
\bauthor{\bsnm{Hyman}, \binits{A.}}:
\batitle{Spindle assembly in xenopus egg extracts: respective roles of centrosomes and microtubule self-organization}.
\bjtitle{J. Cell Biol.}
\bvolume{138}(\bissue{3}),
\bfpage{615}--\blpage{628}
(\byear{1997})
\end{barticle}
\endbibitem

\bibitem[\protect\citeauthoryear{Biswas et~al.}{2021}]{Biswas2021-ph}
\begin{barticle}
\bauthor{\bsnm{Biswas}, \binits{A.}},
\bauthor{\bsnm{Kim}, \binits{K.}},
\bauthor{\bsnm{Cojoc}, \binits{G.}},
\bauthor{\bsnm{Guck}, \binits{J.}},
\bauthor{\bsnm{Reber}, \binits{S.}}:
\batitle{The xenopus spindle is as dense as the surrounding cytoplasm}.
\bjtitle{Dev. Cell}
\bvolume{56}(\bissue{7}),
\bfpage{967}--\blpage{9755}
(\byear{2021})
\end{barticle}
\endbibitem

\bibitem[\protect\citeauthoryear{Desai et~al.}{1998}]{Desai1998-xx}
\begin{bchapter}
\bauthor{\bsnm{Desai}, \binits{A.}},
\bauthor{\bsnm{Murray}, \binits{A.}},
\bauthor{\bsnm{Mitchison}, \binits{T.J.}},
\bauthor{\bsnm{Walczak}, \binits{C.E.}}:
\bctitle{Chapter 20 the use of xenopus egg extracts to study mitotic spindle assembly and function in vitro}.
In: \beditor{\bsnm{Rieder}, \binits{C.L.}} (ed.)
\bbtitle{Methods in Cell Biology}
vol. \bseriesno{61},
pp. \bfpage{385}--\blpage{412}.
\bpublisher{Academic Press}, \blocation{???}
(\byear{1998})
\end{bchapter}
\endbibitem

\bibitem[\protect\citeauthoryear{Gard}{1992}]{Gard1992-qa}
\begin{barticle}
\bauthor{\bsnm{Gard}, \binits{D.L.}}:
\batitle{Microtubule organization during maturation of xenopus oocytes: assembly and rotation of the meiotic spindles}.
\bjtitle{Dev. Biol.}
\bvolume{151}(\bissue{2}),
\bfpage{516}--\blpage{530}
(\byear{1992})
\end{barticle}
\endbibitem

\bibitem[\protect\citeauthoryear{Abercrombie et~al.}{1970}]{Abercrombie1970-id}
\begin{barticle}
\bauthor{\bsnm{Abercrombie}, \binits{M.}},
\bauthor{\bsnm{Heaysman}, \binits{J.E.}},
\bauthor{\bsnm{Pegrum}, \binits{S.M.}}:
\batitle{The locomotion of fibroblasts in culture. {II}. ``{RRuffling}''}.
\bjtitle{Exp. Cell Res.}
\bvolume{60}(\bissue{3}),
\bfpage{437}--\blpage{444}
(\byear{1970})
\end{barticle}
\endbibitem

\bibitem[\protect\citeauthoryear{Li et~al.}{2023}]{Li2023-ev}
\begin{barticle}
\bauthor{\bsnm{Li}, \binits{Y.}},
\bauthor{\bsnm{Ku{\v c}era}, \binits{O.}},
\bauthor{\bsnm{Cuvelier}, \binits{D.}},
\bauthor{\bsnm{Rutkowski}, \binits{D.M.}},
\bauthor{\bsnm{Deygas}, \binits{M.}},
\bauthor{\bsnm{Rai}, \binits{D.}},
\bauthor{\bsnm{Pavlovi{\v c}}, \binits{T.}},
\bauthor{\bsnm{Vicente}, \binits{F.N.}},
\bauthor{\bsnm{Piel}, \binits{M.}},
\bauthor{\bsnm{Giannone}, \binits{G.}},
\bauthor{\bsnm{Vavylonis}, \binits{D.}},
\bauthor{\bsnm{Akhmanova}, \binits{A.}},
\bauthor{\bsnm{Blanchoin}, \binits{L.}},
\bauthor{\bsnm{Th{\'e}ry}, \binits{M.}}:
\batitle{Compressive forces stabilize microtubules in living cells}.
\bjtitle{Nat. Mater.}
\bvolume{22}(\bissue{7}),
\bfpage{913}--\blpage{924}
(\byear{2023})
\end{barticle}
\endbibitem

\bibitem[\protect\citeauthoryear{Theurkauf et~al.}{1992}]{Theurkauf1992-ml}
\begin{barticle}
\bauthor{\bsnm{Theurkauf}, \binits{W.E.}},
\bauthor{\bsnm{Smiley}, \binits{S.}},
\bauthor{\bsnm{Wong}, \binits{M.L.}},
\bauthor{\bsnm{Alberts}, \binits{B.M.}}:
\batitle{Reorganization of the cytoskeleton during drosophila oogenesis: implications for axis specification and intercellular transport}.
\bjtitle{Development}
\bvolume{115}(\bissue{4}),
\bfpage{923}--\blpage{936}
(\byear{1992})
\end{barticle}
\endbibitem

\bibitem[\protect\citeauthoryear{Goldstein and van~de Meent}{2015}]{Goldstein2015-tk}
\begin{barticle}
\bauthor{\bsnm{Goldstein}, \binits{R.E.}},
\bauthor{\bsnm{Meent}, \binits{J.-W.}}:
\batitle{A physical perspective on cytoplasmic streaming}.
\bjtitle{Interface Focus}
\bvolume{5}(\bissue{4}),
\bfpage{20150030}
(\byear{2015})
\end{barticle}
\endbibitem

\bibitem[\protect\citeauthoryear{He et~al.}{2011}]{He2011-re}
\begin{barticle}
\bauthor{\bsnm{He}, \binits{L.}},
\bauthor{\bsnm{Wang}, \binits{X.}},
\bauthor{\bsnm{Montell}, \binits{D.J.}}:
\batitle{Shining light on drosophila oogenesis: live imaging of egg development}.
\bjtitle{Curr. Opin. Genet. Dev.}
\bvolume{21}(\bissue{5}),
\bfpage{612}--\blpage{619}
(\byear{2011})
\end{barticle}
\endbibitem

\bibitem[\protect\citeauthoryear{Bastock and St~Johnston}{2008}]{Bastock2008-mo}
\begin{barticle}
\bauthor{\bsnm{Bastock}, \binits{R.}},
\bauthor{\bsnm{St~Johnston}, \binits{D.}}:
\batitle{Drosophila oogenesis}.
\bjtitle{Curr. Biol.}
\bvolume{18}(\bissue{23}),
\bfpage{1082}--\blpage{7}
(\byear{2008})
\end{barticle}
\endbibitem

\bibitem[\protect\citeauthoryear{Stein et~al.}{2021}]{Stein2021-wv}
\begin{barticle}
\bauthor{\bsnm{Stein}, \binits{D.B.}},
\bauthor{\bsnm{De~Canio}, \binits{G.}},
\bauthor{\bsnm{Lauga}, \binits{E.}},
\bauthor{\bsnm{Shelley}, \binits{M.J.}},
\bauthor{\bsnm{Goldstein}, \binits{R.E.}}:
\batitle{Swirling instability of the microtubule cytoskeleton}.
\bjtitle{Phys. Rev. Lett.}
\bvolume{126}(\bissue{2}),
\bfpage{028103}
(\byear{2021})
\end{barticle}
\endbibitem

\bibitem[\protect\citeauthoryear{Sanchez et~al.}{2012}]{Sanchez2012-jp}
\begin{barticle}
\bauthor{\bsnm{Sanchez}, \binits{T.}},
\bauthor{\bsnm{Chen}, \binits{D.T.N.}},
\bauthor{\bsnm{DeCamp}, \binits{S.J.}},
\bauthor{\bsnm{Heymann}, \binits{M.}},
\bauthor{\bsnm{Dogic}, \binits{Z.}}:
\batitle{Spontaneous motion in hierarchically assembled active matter}.
\bjtitle{Nature}
\bvolume{491}(\bissue{7424}),
\bfpage{431}--\blpage{434}
(\byear{2012})
\end{barticle}
\endbibitem

\bibitem[\protect\citeauthoryear{N{\'e}d{\'e}lec et~al.}{1997}]{Nedelec1997-rw}
\begin{barticle}
\bauthor{\bsnm{N{\'e}d{\'e}lec}, \binits{F.J.}},
\bauthor{\bsnm{Surrey}, \binits{T.}},
\bauthor{\bsnm{Maggs}, \binits{A.C.}},
\bauthor{\bsnm{Leibler}, \binits{S.}}:
\batitle{Self-organization of microtubules and motors}.
\bjtitle{Nature}
\bvolume{389}(\bissue{6648}),
\bfpage{305}--\blpage{308}
(\byear{1997})
\end{barticle}
\endbibitem

\bibitem[\protect\citeauthoryear{Ross et~al.}{2019}]{Ross2019-ol}
\begin{barticle}
\bauthor{\bsnm{Ross}, \binits{T.D.}},
\bauthor{\bsnm{Lee}, \binits{H.J.}},
\bauthor{\bsnm{Qu}, \binits{Z.}},
\bauthor{\bsnm{Banks}, \binits{R.A.}},
\bauthor{\bsnm{Phillips}, \binits{R.}},
\bauthor{\bsnm{Thomson}, \binits{M.}}:
\batitle{Controlling organization and forces in active matter through optically defined boundaries}.
\bjtitle{Nature}
\bvolume{572}(\bissue{7768}),
\bfpage{224}--\blpage{229}
(\byear{2019})
\end{barticle}
\endbibitem

\bibitem[\protect\citeauthoryear{Gardel et~al.}{2008}]{Gardel2008-xz}
\begin{barticle}
\bauthor{\bsnm{Gardel}, \binits{M.L.}},
\bauthor{\bsnm{Kasza}, \binits{K.E.}},
\bauthor{\bsnm{Brangwynne}, \binits{C.P.}},
\bauthor{\bsnm{Liu}, \binits{J.}},
\bauthor{\bsnm{Weitz}, \binits{D.A.}}:
\batitle{Chapter 19: Mechanical response of cytoskeletal networks}.
\bjtitle{Methods Cell Biol.}
\bvolume{89},
\bfpage{487}--\blpage{519}
(\byear{2008})
\end{barticle}
\endbibitem

\bibitem[\protect\citeauthoryear{Needleman et~al.}{2004}]{Needleman2004-fu}
\begin{barticle}
\bauthor{\bsnm{Needleman}, \binits{D.J.}},
\bauthor{\bsnm{Ojeda-Lopez}, \binits{M.A.}},
\bauthor{\bsnm{Raviv}, \binits{U.}},
\bauthor{\bsnm{Miller}, \binits{H.P.}},
\bauthor{\bsnm{Wilson}, \binits{L.}},
\bauthor{\bsnm{Safinya}, \binits{C.R.}}:
\batitle{Higher-order assembly of microtubules by counterions: from hexagonal bundles to living necklaces}.
\bjtitle{Proc. Natl. Acad. Sci. U. S. A.}
\bvolume{101}(\bissue{46}),
\bfpage{16099}--\blpage{16103}
(\byear{2004})
\end{barticle}
\endbibitem

\bibitem[\protect\citeauthoryear{Portran et~al.}{2013}]{Portran2013-da}
\begin{barticle}
\bauthor{\bsnm{Portran}, \binits{D.}},
\bauthor{\bsnm{Zoccoler}, \binits{M.}},
\bauthor{\bsnm{Gaillard}, \binits{J.}},
\bauthor{\bsnm{Stoppin-Mellet}, \binits{V.}},
\bauthor{\bsnm{Neumann}, \binits{E.}},
\bauthor{\bsnm{Arnal}, \binits{I.}},
\bauthor{\bsnm{Martiel}, \binits{J.L.}},
\bauthor{\bsnm{Vantard}, \binits{M.}}:
\batitle{{MAP65/Ase1} promote microtubule flexibility}.
\bjtitle{Mol. Biol. Cell}
\bvolume{24}(\bissue{12}),
\bfpage{1964}--\blpage{1973}
(\byear{2013})
\end{barticle}
\endbibitem

\bibitem[\protect\citeauthoryear{Baas et~al.}{2016}]{Baas2016-kf}
\begin{barticle}
\bauthor{\bsnm{Baas}, \binits{P.W.}},
\bauthor{\bsnm{Rao}, \binits{A.N.}},
\bauthor{\bsnm{Matamoros}, \binits{A.J.}},
\bauthor{\bsnm{Leo}, \binits{L.}}:
\batitle{Stability properties of neuronal microtubules}.
\bjtitle{Cytoskeleton}
\bvolume{73}(\bissue{9}),
\bfpage{442}--\blpage{460}
(\byear{2016})
\end{barticle}
\endbibitem

\bibitem[\protect\citeauthoryear{Toli{\'c}-N{\o}rrelykke}{2008}]{Tolic-Norrelykke2008-jg}
\begin{barticle}
\bauthor{\bsnm{Toli{\'c}-N{\o}rrelykke}, \binits{I.M.}}:
\batitle{Push-me-pull-you: how microtubules organize the cell interior}.
\bjtitle{Eur. Biophys. J.}
\bvolume{37}(\bissue{7}),
\bfpage{1271}--\blpage{1278}
(\byear{2008})
\end{barticle}
\endbibitem

\bibitem[\protect\citeauthoryear{Kapitein and Hoogenraad}{2015}]{Kapitein2015-sm}
\begin{barticle}
\bauthor{\bsnm{Kapitein}, \binits{L.C.}},
\bauthor{\bsnm{Hoogenraad}, \binits{C.C.}}:
\batitle{Building the neuronal microtubule cytoskeleton}.
\bjtitle{Neuron}
\bvolume{87}(\bissue{3}),
\bfpage{492}--\blpage{506}
(\byear{2015})
\end{barticle}
\endbibitem

\bibitem[\protect\citeauthoryear{Forth and Kapoor}{2017}]{Forth2017-mv}
\begin{barticle}
\bauthor{\bsnm{Forth}, \binits{S.}},
\bauthor{\bsnm{Kapoor}, \binits{T.M.}}:
\batitle{The mechanics of microtubule networks in cell division}.
\bjtitle{J. Cell Biol.}
\bvolume{216}(\bissue{6}),
\bfpage{1525}--\blpage{1531}
(\byear{2017})
\end{barticle}
\endbibitem

\bibitem[\protect\citeauthoryear{Yang et~al.}{2022}]{Yang2022-uh}
\begin{botherref}
\oauthor{\bsnm{Yang}, \binits{F.}},
\oauthor{\bsnm{Liu}, \binits{S.}},
\oauthor{\bsnm{Lee}, \binits{H.J.}},
\oauthor{\bsnm{Phillips}, \binits{R.}},
\oauthor{\bsnm{Thomson}, \binits{M.}}:
Dynamic flow control through active matter programming language
(2022)
{\href{https://arxiv.org/abs/2208.12839}{{arXiv:2208.12839}}}
{[cond-mat.soft]}
\end{botherref}
\endbibitem

\bibitem[\protect\citeauthoryear{Brugu{\'e}s and Needleman}{2014}]{Brugues2014-kc}
\begin{barticle}
\bauthor{\bsnm{Brugu{\'e}s}, \binits{J.}},
\bauthor{\bsnm{Needleman}, \binits{D.}}:
\batitle{Physical basis of spindle self-organization}.
\bjtitle{Proceedings of the National Academy of Sciences}
\bvolume{111}(\bissue{52}),
\bfpage{18496}--\blpage{18500}
(\byear{2014})
\end{barticle}
\endbibitem

\bibitem[\protect\citeauthoryear{Sanchez et~al.}{2005}]{Sanchez2005-tb}
\begin{barticle}
\bauthor{\bsnm{Sanchez}, \binits{C.}},
\bauthor{\bsnm{Arribart}, \binits{H.}},
\bauthor{\bsnm{Guille}, \binits{M.M.G.}}:
\batitle{Biomimetism and bioinspiration as tools for the design of innovative materials and systems}.
\bjtitle{Nat. Mater.}
\bvolume{4}(\bissue{4}),
\bfpage{277}--\blpage{288}
(\byear{2005})
\end{barticle}
\endbibitem

\bibitem[\protect\citeauthoryear{Brangwynne et~al.}{2006}]{Brangwynne2006-my}
\begin{barticle}
\bauthor{\bsnm{Brangwynne}, \binits{C.P.}},
\bauthor{\bsnm{MacKintosh}, \binits{F.C.}},
\bauthor{\bsnm{Kumar}, \binits{S.}},
\bauthor{\bsnm{Geisse}, \binits{N.A.}},
\bauthor{\bsnm{Talbot}, \binits{J.}},
\bauthor{\bsnm{Mahadevan}, \binits{L.}},
\bauthor{\bsnm{Parker}, \binits{K.K.}},
\bauthor{\bsnm{Ingber}, \binits{D.E.}},
\bauthor{\bsnm{Weitz}, \binits{D.A.}}:
\batitle{Microtubules can bear enhanced compressive loads in living cells because of lateral reinforcement}.
\bjtitle{J. Cell Biol.}
\bvolume{173}(\bissue{5}),
\bfpage{733}--\blpage{741}
(\byear{2006})
\end{barticle}
\endbibitem

\bibitem[\protect\citeauthoryear{Whitesides}{2015}]{Whitesides2015-ca}
\begin{barticle}
\bauthor{\bsnm{Whitesides}, \binits{G.M.}}:
\batitle{Bioinspiration: something for everyone}.
\bjtitle{Interface Focus}
\bvolume{5}(\bissue{4}),
\bfpage{20150031}
(\byear{2015})
\end{barticle}
\endbibitem

\bibitem[\protect\citeauthoryear{Siegrist and Doe}{2007}]{Siegrist2007-xl}
\begin{barticle}
\bauthor{\bsnm{Siegrist}, \binits{S.E.}},
\bauthor{\bsnm{Doe}, \binits{C.Q.}}:
\batitle{Microtubule-induced cortical cell polarity}.
\bjtitle{Genes Dev.}
\bvolume{21}(\bissue{5}),
\bfpage{483}--\blpage{496}
(\byear{2007})
\end{barticle}
\endbibitem

\bibitem[\protect\citeauthoryear{Dogterom and Surrey}{2013}]{Dogterom2013-ko}
\begin{barticle}
\bauthor{\bsnm{Dogterom}, \binits{M.}},
\bauthor{\bsnm{Surrey}, \binits{T.}}:
\batitle{Microtubule organization in vitro}.
\bjtitle{Curr. Opin. Cell Biol.}
\bvolume{25}(\bissue{1}),
\bfpage{23}--\blpage{29}
(\byear{2013})
\end{barticle}
\endbibitem

\bibitem[\protect\citeauthoryear{Petry et~al.}{2013}]{Petry2013-il}
\begin{barticle}
\bauthor{\bsnm{Petry}, \binits{S.}},
\bauthor{\bsnm{Groen}, \binits{A.C.}},
\bauthor{\bsnm{Ishihara}, \binits{K.}},
\bauthor{\bsnm{Mitchison}, \binits{T.J.}},
\bauthor{\bsnm{Vale}, \binits{R.D.}}:
\batitle{Branching microtubule nucleation in xenopus egg extracts mediated by augmin and {TPX2}}.
\bjtitle{Cell}
\bvolume{152}(\bissue{4}),
\bfpage{768}--\blpage{777}
(\byear{2013})
\end{barticle}
\endbibitem

\bibitem[\protect\citeauthoryear{Brugu{\'e}s et~al.}{2012}]{Brugues2012-qi}
\begin{barticle}
\bauthor{\bsnm{Brugu{\'e}s}, \binits{J.}},
\bauthor{\bsnm{Nuzzo}, \binits{V.}},
\bauthor{\bsnm{Mazur}, \binits{E.}},
\bauthor{\bsnm{Needleman}, \binits{D.J.}}:
\batitle{Nucleation and transport organize microtubules in metaphase spindles}.
\bjtitle{Cell}
\bvolume{149}(\bissue{3}),
\bfpage{554}--\blpage{564}
(\byear{2012})
\end{barticle}
\endbibitem

\bibitem[\protect\citeauthoryear{Kruse et~al.}{2004}]{Kruse2004-vq}
\begin{barticle}
\bauthor{\bsnm{Kruse}, \binits{K.}},
\bauthor{\bsnm{Joanny}, \binits{J.F.}},
\bauthor{\bsnm{J{\"u}licher}, \binits{F.}},
\bauthor{\bsnm{Prost}, \binits{J.}},
\bauthor{\bsnm{Sekimoto}, \binits{K.}}:
\batitle{Asters, vortices, and rotating spirals in active gels of polar filaments}.
\bjtitle{Phys. Rev. Lett.}
\bvolume{92}(\bissue{7}),
\bfpage{078101}
(\byear{2004})
\end{barticle}
\endbibitem

\bibitem[\protect\citeauthoryear{Hentrich and Surrey}{2010}]{Hentrich2010-vp}
\begin{barticle}
\bauthor{\bsnm{Hentrich}, \binits{C.}},
\bauthor{\bsnm{Surrey}, \binits{T.}}:
\batitle{Microtubule organization by the antagonistic mitotic motors kinesin-5 and kinesin-14}.
\bjtitle{J. Cell Biol.}
\bvolume{189}(\bissue{3}),
\bfpage{465}--\blpage{480}
(\byear{2010})
\end{barticle}
\endbibitem

\bibitem[\protect\citeauthoryear{Doostmohammadi et~al.}{2018}]{Doostmohammadi2018-td}
\begin{barticle}
\bauthor{\bsnm{Doostmohammadi}, \binits{A.}},
\bauthor{\bsnm{Ign{\'e}s-Mullol}, \binits{J.}},
\bauthor{\bsnm{Yeomans}, \binits{J.M.}},
\bauthor{\bsnm{Sagu{\'e}s}, \binits{F.}}:
\batitle{Active nematics}.
\bjtitle{Nat. Commun.}
\bvolume{9}(\bissue{1}),
\bfpage{3246}
(\byear{2018})
\end{barticle}
\endbibitem

\bibitem[\protect\citeauthoryear{Chen}{2019}]{Chen2019-ke}
\begin{barticle}
\bauthor{\bsnm{Chen}, \binits{J.}}:
\batitle{Two-point statistics of coherent structure in turbulent flow}.
\bjtitle{J. Flow Control Meas. Amp Vis.}
\bvolume{07}(\bissue{04}),
\bfpage{153}--\blpage{173}
(\byear{2019})
\end{barticle}
\endbibitem

\bibitem[\protect\citeauthoryear{Najma et~al.}{2023}]{Najma2023-ya}
\begin{botherref}
\oauthor{\bsnm{Najma}, \binits{B.}},
\oauthor{\bsnm{Baskaran}, \binits{A.}},
\oauthor{\bsnm{Foster}, \binits{P.J.}},
\oauthor{\bsnm{Duclos}, \binits{G.}}:
Microscopic interactions control a structural transition in active mixtures of microtubules and molecular motors
(2023)
\end{botherref}
\endbibitem

\bibitem[\protect\citeauthoryear{Nedelec and Foethke}{2007}]{Nedelec2007-ln}
\begin{barticle}
\bauthor{\bsnm{Nedelec}, \binits{F.}},
\bauthor{\bsnm{Foethke}, \binits{D.}}:
\batitle{Collective langevin dynamics of flexible cytoskeletal fibers}.
\bjtitle{New J. Phys.}
\bvolume{9}(\bissue{11}),
\bfpage{427}
(\byear{2007})
\end{barticle}
\endbibitem

\bibitem[\protect\citeauthoryear{Odde et~al.}{1995}]{Odde1995-gb}
\begin{barticle}
\bauthor{\bsnm{Odde}, \binits{D.J.}},
\bauthor{\bsnm{Cassimeris}, \binits{L.}},
\bauthor{\bsnm{Buettner}, \binits{H.M.}}:
\batitle{Kinetics of microtubule catastrophe assessed by probabilistic analysis}.
\bjtitle{Biophys. J.}
\bvolume{69}(\bissue{3}),
\bfpage{796}--\blpage{802}
(\byear{1995})
\end{barticle}
\endbibitem

\bibitem[\protect\citeauthoryear{Artime and De~Domenico}{2021}]{Artime2021-iv}
\begin{barticle}
\bauthor{\bsnm{Artime}, \binits{O.}},
\bauthor{\bsnm{De~Domenico}, \binits{M.}}:
\batitle{Percolation on feature-enriched interconnected systems}.
\bjtitle{Nat. Commun.}
\bvolume{12}(\bissue{1}),
\bfpage{2478}
(\byear{2021})
\end{barticle}
\endbibitem

\bibitem[\protect\citeauthoryear{Surrey et~al.}{2001}]{Surrey2001-nl}
\begin{barticle}
\bauthor{\bsnm{Surrey}, \binits{T.}},
\bauthor{\bsnm{N{\'e}d{\'e}lec}, \binits{F.}},
\bauthor{\bsnm{Leibler}, \binits{S.}},
\bauthor{\bsnm{Karsenti}, \binits{E.}}:
\batitle{Physical properties determining self-organization of motors and microtubules}.
\bjtitle{Science}
\bvolume{292}(\bissue{5519}),
\bfpage{1167}--\blpage{1171}
(\byear{2001})
\end{barticle}
\endbibitem

\bibitem[\protect\citeauthoryear{Xie and Jian}{2020}]{Xie2020-jc}
\begin{barticle}
\bauthor{\bsnm{Xie}, \binits{Z.}},
\bauthor{\bsnm{Jian}, \binits{Y.}}:
\batitle{Electrokinetic energy conversion of nanofluids in {MHD-based} microtube}.
\bjtitle{Energy}
\bvolume{212},
\bfpage{118711}
(\byear{2020})
\end{barticle}
\endbibitem

\bibitem[\protect\citeauthoryear{Eriksson et~al.}{2007}]{Eriksson2007-og}
\begin{barticle}
\bauthor{\bsnm{Eriksson}, \binits{E.}},
\bauthor{\bsnm{Enger}, \binits{J.}},
\bauthor{\bsnm{Nordlander}, \binits{B.}},
\bauthor{\bsnm{Erjavec}, \binits{N.}},
\bauthor{\bsnm{Ramser}, \binits{K.}},
\bauthor{\bsnm{Goks{\"o}r}, \binits{M.}},
\bauthor{\bsnm{Hohmann}, \binits{S.}},
\bauthor{\bsnm{Nystr{\"o}m}, \binits{T.}},
\bauthor{\bsnm{Hanstorp}, \binits{D.}}:
\batitle{A microfluidic system in combination with optical tweezers for analyzing rapid and reversible cytological alterations in single cells upon environmental changes}.
\bjtitle{Lab Chip}
\bvolume{7}(\bissue{1}),
\bfpage{71}--\blpage{76}
(\byear{2007})
\end{barticle}
\endbibitem

\end{thebibliography}

\end{document}


\maketitle

\section{PIV and Optical Flow Methods}

\subsection{Optical Flow}
Optical flow is a method used to estimate the motion of objects between consecutive frames of a video sequence by analyzing the apparent motion of brightness patterns in the images. The Farneback method is a specific optical flow algorithm that uses polynomial expansions to estimate the velocity field.

Code implementation in optical\textunderscore flow.ipynb

\subsubsection{Optical Flow Assumptions}
The basic assumption of optical flow is the brightness constancy constraint, which states that the intensity of a pixel remains constant between consecutive frames. Mathematically, this can be expressed as:
\begin{equation}
I_1(\mathbf{x}) = I_2(\mathbf{x} + \mathbf{d}(\mathbf{x})),
\end{equation}
where $I_1$ and $I_2$ are the intensities of the pixel at position $\mathbf{x}$ in the two frames, and $\mathbf{d}(\mathbf{x})$ is the velocity vector of the pixel.

\subsubsection{Farneback Method}
The Farneback method approximates the velocity field using polynomial expansions. It represents the local neighborhood of each pixel by a polynomial, which can be expressed as:
\begin{equation}
I(\mathbf{x}) \approx \mathbf{a}_0 + \mathbf{a}_1 x + \mathbf{a}_2 y + \mathbf{a}_3 x^2 + \mathbf{a}_4 xy + \mathbf{a}_5 y^2,
\end{equation}
where $\mathbf{a}_i$ are the coefficients of the polynomial.

To estimate the velocity field, the method constructs an over-determined system of linear equations by sampling the polynomial coefficients in a neighborhood around each pixel. This system is then solved using least squares minimization to find the velocity vector $\mathbf{d}(\mathbf{x})$ that best fits the brightness constancy constraint.

\subsubsection{Velocity Field Calculation}
Once the velocity field $\mathbf{d}(\mathbf{x})$ is obtained, the velocity field $\mathbf{v}(\mathbf{x})$ is calculated by dividing the displacement by the time interval $\Delta t$ between frames:
\begin{equation}
\mathbf{v}(\mathbf{x}) = \frac{\mathbf{d}(\mathbf{x})}{\Delta t}.
\end{equation}

\subsection{Particle Image Velocimetry (PIV)}
Particle Image Velocimetry (PIV) is a widely used technique for measuring fluid flow velocity fields by tracking the displacement of tracer particles between sequential image frames. This method provides a non-intrusive way to capture detailed flow dynamics over an entire plane, making it ideal for a variety of fluid dynamics studies.

\subsubsection{PIV Processing Parameters}
In our analysis, PIV was conducted using a three-pass algorithm to enhance the accuracy and resolution of the velocity field. Each pass refines the previous estimate by using smaller interrogation windows and step sizes, effectively increasing spatial resolution. The parameters used for each pass are:

\begin{itemize}
    \item First pass: Interrogation window of 64 pixels, step size of 32 pixels.
    \item Second pass: Interrogation window of 32 pixels, step size of 16 pixels.
    \item Third pass: Interrogation window of 16 pixels, step size of 8 pixels.
\end{itemize}

These parameters are chosen to balance the trade-off between spatial resolution and robustness against noise.

\subsubsection{Sub-Pixel Estimation and Correlation}
For sub-pixel accuracy in the displacement measurement, we employed the Gauss 2x3-point estimator. This method fits a Gaussian function to the cross-correlation peak to estimate the displacement vector with sub-pixel precision. The displacement $\mathbf{d}(\mathbf{x})$ can be expressed as:

\begin{equation}
\mathbf{d}(\mathbf{x}) = \frac{\partial \Phi}{\partial \mathbf{x}} \bigg/ \frac{\partial^2 \Phi}{\partial \mathbf{x}^2},
\end{equation}

where $\Phi$ represents the cross-correlation function. The peak of this function is iteratively refined to achieve higher precision.

\subsubsection{Post-Processing and Validation}
To ensure the reliability of the PIV results, post-processing steps were applied. Vector validation routines, including outlier detection and replacement, were performed using a threshold of 8 times the standard deviation. Additionally, a local median threshold of 3 was applied to detect and correct erroneous vectors. This validation step is crucial for eliminating spurious vectors and improving the overall quality of the velocity field.

\subsubsection{PIV Velocity Field Calculation}
The final velocity field $\mathbf{v}(\mathbf{x})$ is calculated by dividing the measured displacement $\mathbf{d}(\mathbf{x})$ by the time interval $\Delta t$ between frames:

\begin{equation}
\mathbf{v}(\mathbf{x}) = \frac{\mathbf{d}(\mathbf{x})}{\Delta t}.
\end{equation}

This velocity field provides a detailed representation of the flow dynamics at each interrogation window's center.

\subsection{Grid Averaging and Method Comparison}
For comparative analysis between PIV and optical flow, both velocity fields need to be averaged over a uniform grid to facilitate direct comparison.

\subsubsection{Grid Averaging Process}
The grid averaging process involves interpolating the PIV and optical flow velocity data onto a common grid. Given a grid size \( G \), the flow field is averaged over each grid cell. The velocity components \( u \) and \( v \) are computed as:

\begin{equation}
u_{avg}(i, j) = \frac{1}{n} \sum_{k=1}^{n} u_k,
\end{equation}

\begin{equation}
v_{avg}(i, j) = \frac{1}{n} \sum_{k=1}^{n} v_k,
\end{equation}

where \( u_k \) and \( v_k \) are the velocity components within the grid cell \((i, j)\), and \( n \) is the number of vectors within the cell. This averaging smooths the velocity field, reducing noise and allowing for a clearer comparison of the flow structures between methods.

\subsubsection{Comparative Analysis}

To compare the performance of the PIV and optical flow methods, we calculate three key metrics: relative speed errors, relative orientation errors, and cross-correlation between the methods and ground truth.

\paragraph{Relative Speed Error}

The relative speed error quantifies the discrepancy in the magnitudes of the velocity vectors between the method under evaluation and the ground truth. It is defined as:

\begin{equation}
\text{Relative Speed Error} = \frac{|V_{\text{method}} - V_{\text{gt}}|}{V_{\text{gt}}},
\end{equation}

where \( V_{\text{method}} \) and \( V_{\text{gt}} \) are the magnitudes of the velocity vectors from the method under consideration (either PIV or optical flow) and the ground truth, respectively. This metric highlights the magnitude difference, providing insight into how accurately each method estimates the speed of the flow.

\paragraph{Relative Orientation Error}

The relative orientation error assesses the angular deviation between the estimated velocity vector from the method and the ground truth vector. It is calculated as follows:

\begin{equation}
\text{Relative Orientation Error} = \cos^{-1} \left( \frac{\mathbf{v}_{\text{method}} \cdot \mathbf{v}_{\text{gt}}}{|\mathbf{v}_{\text{method}}||\mathbf{v}_{\text{gt}}|} \right),
\end{equation}

where \( \mathbf{v}_{\text{method}} \) and \( \mathbf{v}_{\text{gt}} \) are the velocity vectors obtained from the method and the ground truth, respectively. This metric measures the angular difference between the vectors, indicating how well each method captures the direction of the flow.

\paragraph{Cross-Correlation Analysis}

To evaluate the temporal consistency and overall agreement of the methods with the ground truth, we compute the cross-correlation of the mean speed across frames. The cross-correlation between the method's mean speed and the ground truth's mean speed is defined as:

\begin{equation}
\text{Cross-Correlation} = \frac{\sum (V_{\text{method}} - \bar{V}_{\text{method}})(V_{\text{gt}} - \bar{V}_{\text{gt}})}{\sqrt{\sum (V_{\text{method}} - \bar{V}_{\text{method}})^2 \sum (V_{\text{gt}} - \bar{V}_{\text{gt}})^2}},
\end{equation}

where \( V_{\text{method}} \) and \( V_{\text{gt}} \) are the mean speeds of the method and ground truth for a given frame, and \( \bar{V}_{\text{method}} \) and \( \bar{V}_{\text{gt}} \) are the overall mean speeds across all frames. This metric assesses how well the trends in the method's velocity estimates correlate with those of the ground truth over time.

\subsubsection{Benchmarking}
We benchmarked our custom optical flow algorithm with openPIV package from python. We used experimental data from the our previous work \cite{Yang2022-uh}. The dataset consist of  densely labeled Alexa-647 MTs and 1 µm tracer beads were sequentially imaged every 8 seconds using fluorescence filter or brightfield microscopy. We performed PIV on the tracer beads data as the ground truth and performed PIV and image flow on the densely labeled data. We noticed that our image flow algorithm performed better in calculating both the magnitude of the velocity field and the orientation of the velocity vectors as shown in supplemental figure 1.
\begin{figure}[h]
\centering
\includegraphics[width=0.8\textwidth]{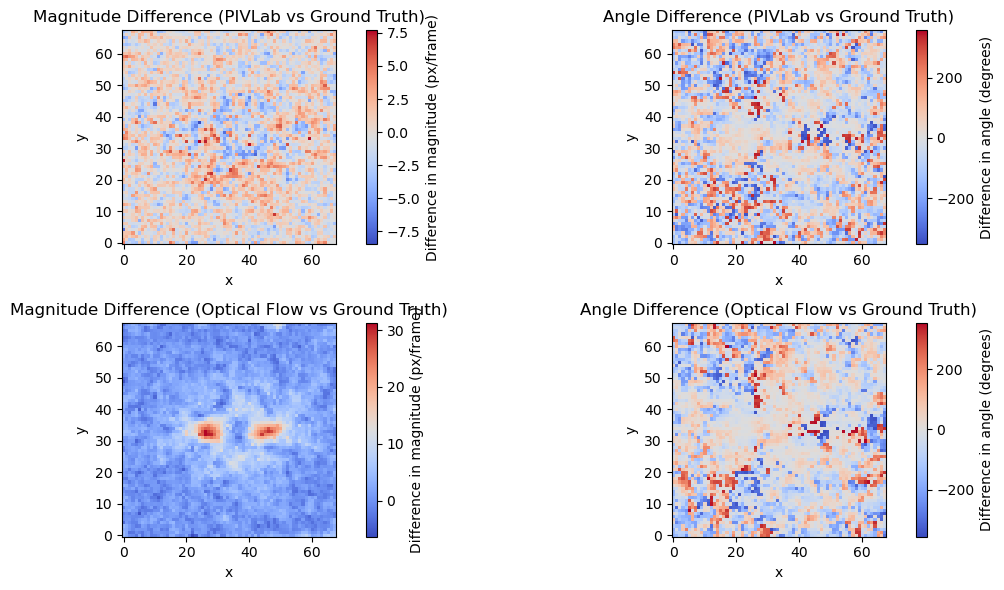}
\caption{
\textbf{Comparison of Magnitude and Angle Differences between PIV/Optical Flow and Ground Truth for Frame 20.} 
The top row shows the \textbf{PIVLab} method compared to the ground truth: (left) magnitude difference and (right) angle difference. The bottom row shows the \textbf{Optical Flow} method compared to the ground truth: (left) magnitude difference and (right) angle difference. The magnitude difference is measured in pixels per frame, and the angle difference is in degrees. The PIVLab results exhibit less pronounced differences in magnitude and a scattered pattern of angle differences. In contrast, the optical flow method shows localized areas with significant magnitude differences but more uniform angle discrepancies.
}
\end{figure}

\begin{figure}[h]
\centering
\includegraphics[width=0.8\textwidth]{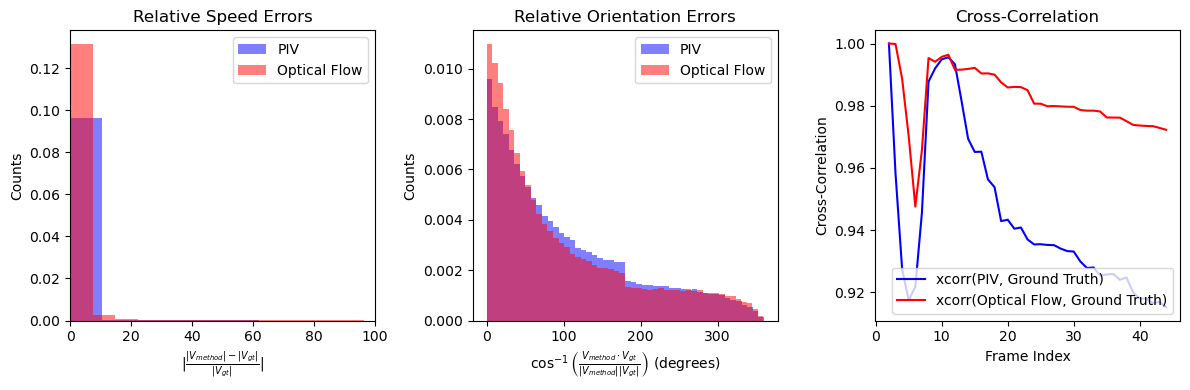}
\caption{
\textbf{Histograms of Relative Speed and Orientation Errors, and Cross-Correlation for PIV and Optical Flow Methods.} 
From left to right: 
(1) The \textbf{relative speed errors} histogram shows the distribution of speed discrepancies normalized by the ground truth velocity.
(2) The \textbf{relative orientation errors} histogram shows the distribution of angular discrepancies between the estimated and ground truth vectors.
(3) The \textbf{cross-correlation} plot between the methods' mean speed and the ground truth's mean speed over 45 frames. Higher cross-correlation values indicate better temporal consistency with the ground truth. The optical flow method (red) maintains a high and stable correlation, while the PIV method (blue) shows a decline in correlation over time.
}
\end{figure}

\clearpage

\section{Microtubule Concentration Calibration}
To establish a quantitative relationship between microtubule concentration and image intensity, we developed a calibration curve using a controlled experimental setup. This calibration allows us to convert observed fluorescence intensities in our main experiments to  microtubule densities.

Code implementation in figure\textunderscore 1 \textunderscore density.ipynb

\subsection{Experimental Procedure}

We prepared microtubule solutions at eight different concentrations.
Each concentration was loaded into a separate flow cell.
For each concentration, we captured 10 fluorescence microscopy images at different locations within the flow cell.

\subsection{Data Analysis}

For each image, we calculated the average pixel intensity.
We then computed the mean intensity and standard deviation across the 10 images for each concentration.
The known microtubule concentrations were converted to densities (MT/$\mu$m²) based on the flow cell dimensions.

\subsection{Calibration Curve}
We plotted the average pixel intensity against the microtubule density and performed a linear regression. The resulting calibration equation is:
\begin{equation}
I = 143.05\rho - 39.76
\end{equation}
where:
\begin{itemize}
\item $I$ is the average pixel intensity
\item $\rho$ is the microtubule density in MT/$\mu$m²
\end{itemize}

\begin{figure}[h]
\centering
\includegraphics[width=0.8\textwidth]{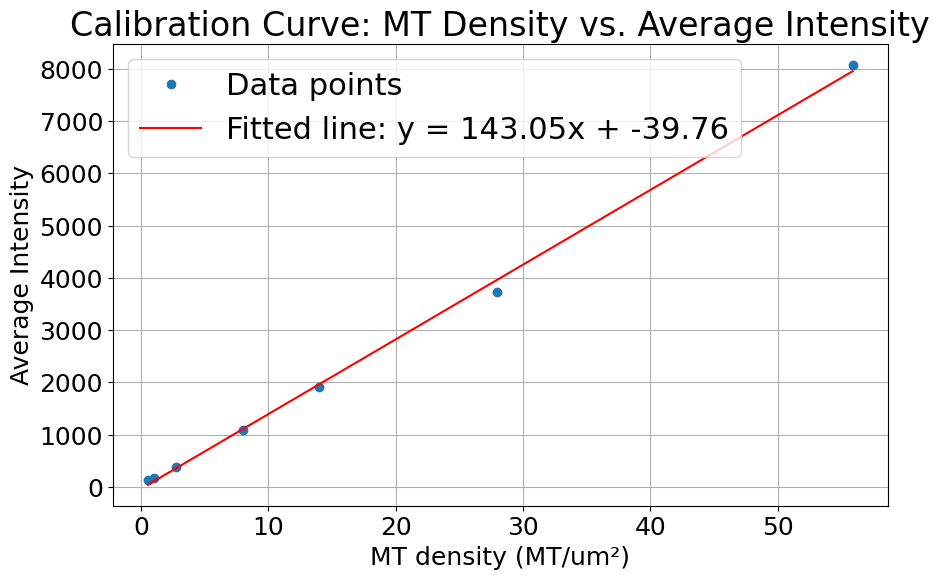}
\caption{
\textbf{Microtubule density calibration curve} }
\end{figure}

\section{Velocity Correlation Function and Correlation Length Calculation}
To quantify the spatial coherence of the velocity field in our active microtubule network, we computed the two-point velocity correlation function and derived the correlation length. This analysis provides insights into the length scale over which the system exhibits coordinated motion.

Code implementation in figure\textunderscore 1 \textunderscore correlation.ipynb
\subsection{Velocity Correlation Function}
The two-point velocity correlation function $C(r)$ is defined as:
\begin{equation}
C(r) = \frac{\langle \mathbf{v}(\mathbf{r}_0) \cdot \mathbf{v}(\mathbf{r}_0 + \mathbf{r}) \rangle}{\langle |\mathbf{v}(\mathbf{r}_0)|^2 \rangle}
\end{equation}
where $\mathbf{v}(\mathbf{r})$ is the velocity vector at position $\mathbf{r}$, and the angle brackets denote an ensemble average over all positions $\mathbf{r}_0$ and time. Velocity vectors were computed based on Supplemental Information 1.1
In practice, we compute this correlation function as follows:

We select a region of interest in the center of the image, spanning ±100 pixels in the y-direction.
We compute the mean velocity field by averaging over frames 90 to 110.
For each frame, we calculate the velocity fluctuations: $\delta \mathbf{v}(\mathbf{r}) = \mathbf{v}(\mathbf{r}) - \langle \mathbf{v}(\mathbf{r}) \rangle$
We then compute the correlation between a reference point (rightmost column of the ROI) and all other points to its left:

\begin{equation}
C(r) = \frac{\langle \delta \mathbf{v}(\mathbf{r}_\text{ref}) \cdot \delta \mathbf{v}(\mathbf{r}_\text{ref} + \mathbf{r}) \rangle}{\sqrt{\langle |\delta \mathbf{v}(\mathbf{r}_\text{ref})|^2 \rangle \langle |\delta \mathbf{v}(\mathbf{r}_\text{ref} + \mathbf{r})|^2 \rangle}}
\end{equation}

We average this correlation over all frames to obtain the final correlation function.

\subsection{Correlation Length Calculation}
The correlation length $L_c$ is a measure of the distance over which velocities remain correlated. We use two methods to calculate $L_c$, depending on the behavior of the correlation function:

For short incubation times (local phase):
We fit the correlation function to an exponential decay:

\begin{equation}
C(r) = A e^{-r/L_c}
\end{equation}
where $A$ is the amplitude and $L_c$ is the correlation length. We use non-linear least squares fitting to determine $L_c$.

For long incubation times (global phase):
The correlation function exhibits more complex behavior, often becoming negative at larger distances. In this case, we define $L_c$ as the distance at which the correlation function first drops to $1/e$ of its initial value:

\begin{equation}
C(L_c) = \frac{1}{e}C(0)
\end{equation}

\begin{figure}[h]
\centering
\includegraphics[width=1\textwidth]{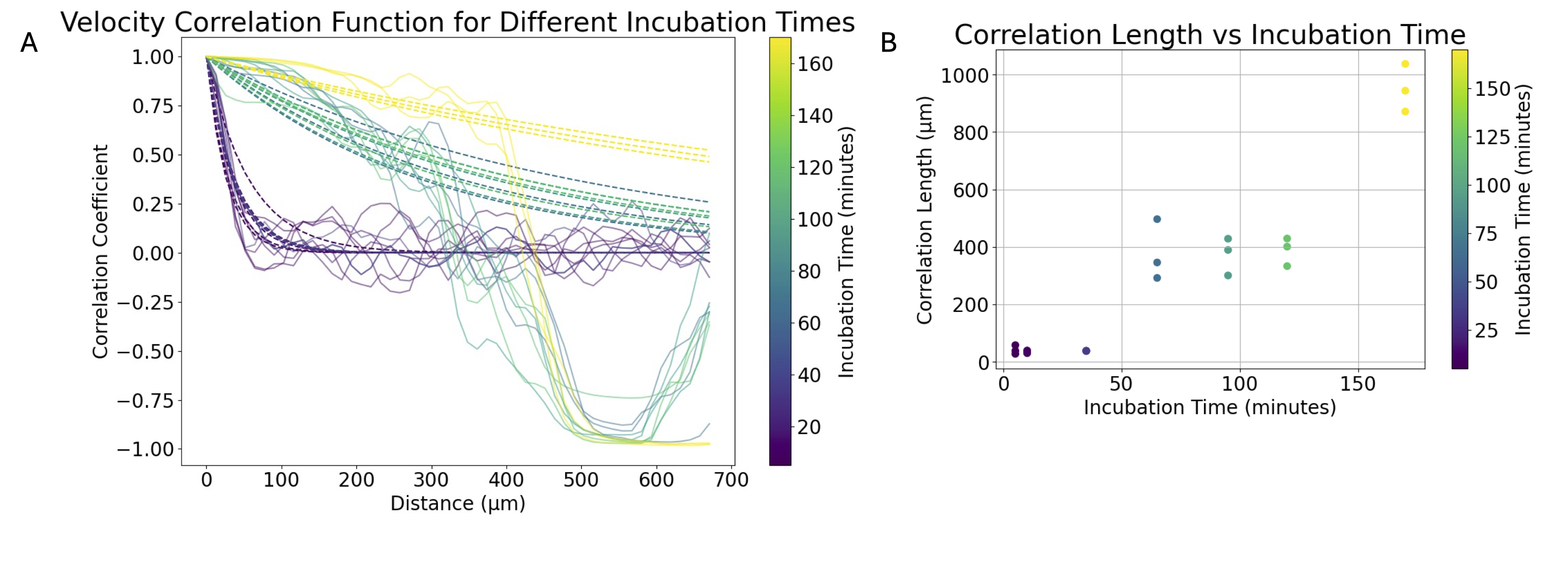}
\caption{
\textbf{Exponential decay fitted to the two-point correlation function, and the corresponding correlation length l}}
\end{figure}

\section{Flux Analysis}
Code implementation in figure\textunderscore 1 \textunderscore flux.ipynb

\subsection{Elliptical Region of Interest (ROI) Definition}
We define an elliptical ROI using the equation:

\begin{equation}
    \left(\frac{x - x_c}{a}\right)^2 + \left(\frac{y - y_c}{b}\right)^2 \leq 1
\end{equation}

where $(x_c, y_c)$ is the center of the ellipse, and $a$ and $b$ are the semi-major and semi-minor axes, respectively.

\subsection{Boundary Point Selection}
We use a k-d tree algorithm to efficiently find the nearest points on the ROI boundary to a set of evenly distributed points on a perfect ellipse. This ensures uniform sampling around the ROI while adhering to the pixelated nature of the image.

\subsection{Flux Calculation}
For each boundary point $i$ at time $t$, we calculate the flux $F_i(t)$ as:

\begin{equation}
    F_i(t) = \pm v_x(i,t) \cdot \rho(i,t) \cdot \Delta x
\end{equation}

where $v_x(i,t)$ is the x-component of the velocity, $\rho(i,t)$ is the microtubule density, and $\Delta x$ is the pixel size. The sign is positive for points on the left half of the ellipse and negative for points on the right half.

\subsection{Density Calibration}
We convert intensity to microtubule density using the linear relationship using the equation obtained in section 2.

\subsection{Data Smoothing}
We apply Gaussian smoothing to the flux data:

\begin{equation}
    F_\text{smooth}(x) = \int F(t) \cdot G(x-t) dt
\end{equation}

where $G(x)$ is a Gaussian function with standard deviation $\sigma = 2$.

\subsection{Cumulative and Total Flux}
Cumulative flux for each point $i$: $C_i = \sum_t F_i(t)$
Total flux at time $t$: $T(t) = \sum_i F_i(t)$

\subsection{Velocity Field Interpolation}
We interpolate the velocity field to the ROI boundary points using a grid-based approach:

\begin{equation}
    v(x,y) \approx v(\lfloor x/g \rfloor \cdot g, \lfloor y/g \rfloor \cdot g)
\end{equation}

where $g$ is the grid size (30 pixels in this case).
\begin{figure}[h]
\centering
\includegraphics[width=1\textwidth]{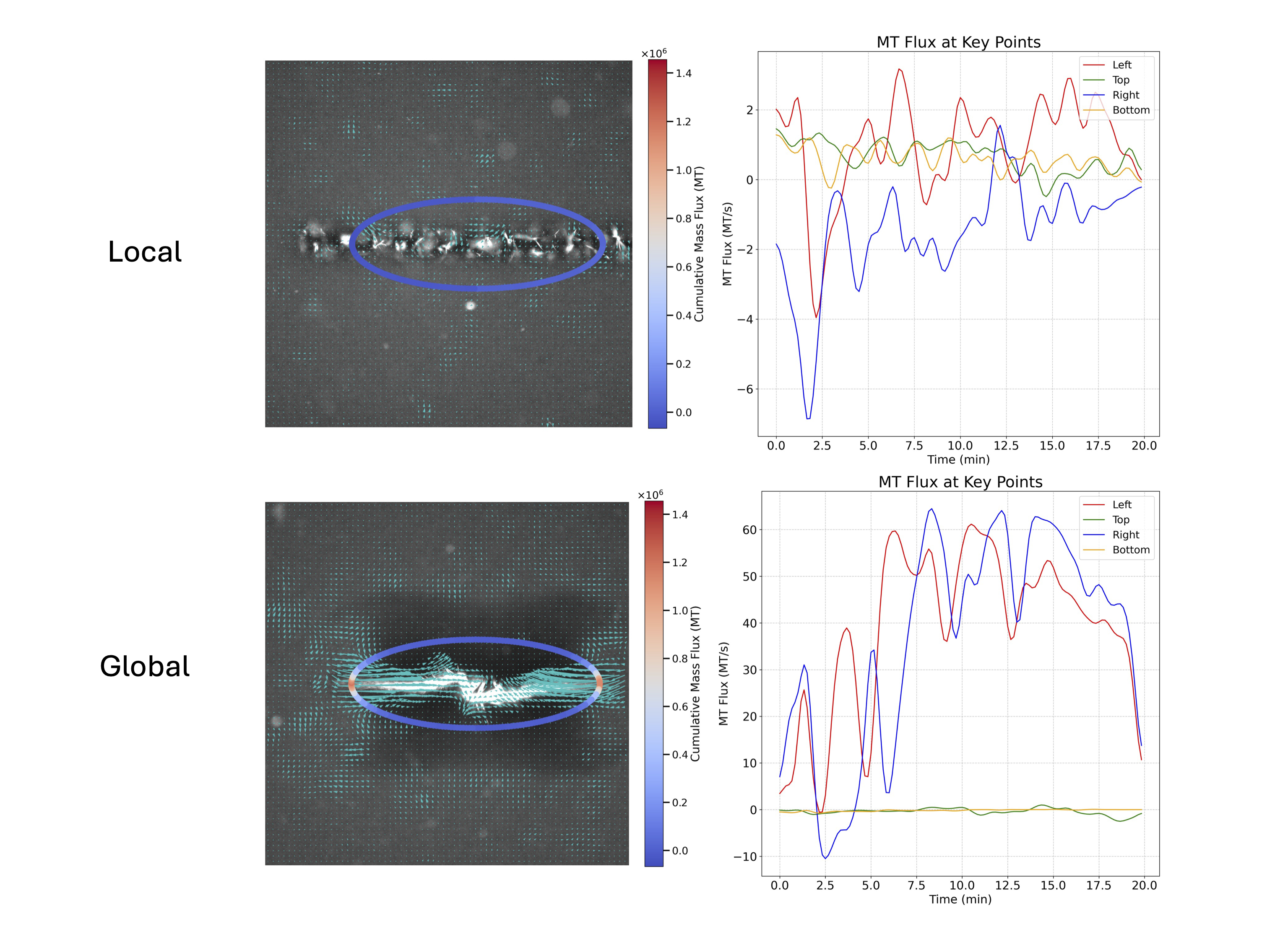}
\caption{
\textbf{Microtubule flux at vertices and co-vertices of the ROI in representative global and local phase. }}
\end{figure}

\section{Fluorescence Recovery After Photobleaching (FRAP)}

Fluorescence Recovery After Photobleaching (FRAP) is a technique used to study the dynamics of molecular diffusion and interactions within cellular environments. In a FRAP experiment, a region of interest (ROI) within a fluorescently labeled sample is photobleached using a high-intensity laser, and the subsequent recovery of fluorescence within the bleached area is monitored over time. The rate of fluorescence recovery provides insights into the mobility and binding characteristics of the molecules in the ROI. Since we cannot directly image and measure the microtubule structures in the experiments, we employed FRAP to measure for the microtubule structure size.
Code implementation in figure\textunderscore 2 \textunderscore FRAP.ipynb

\begin{figure}[h]
\centering
\includegraphics[width=1\textwidth]{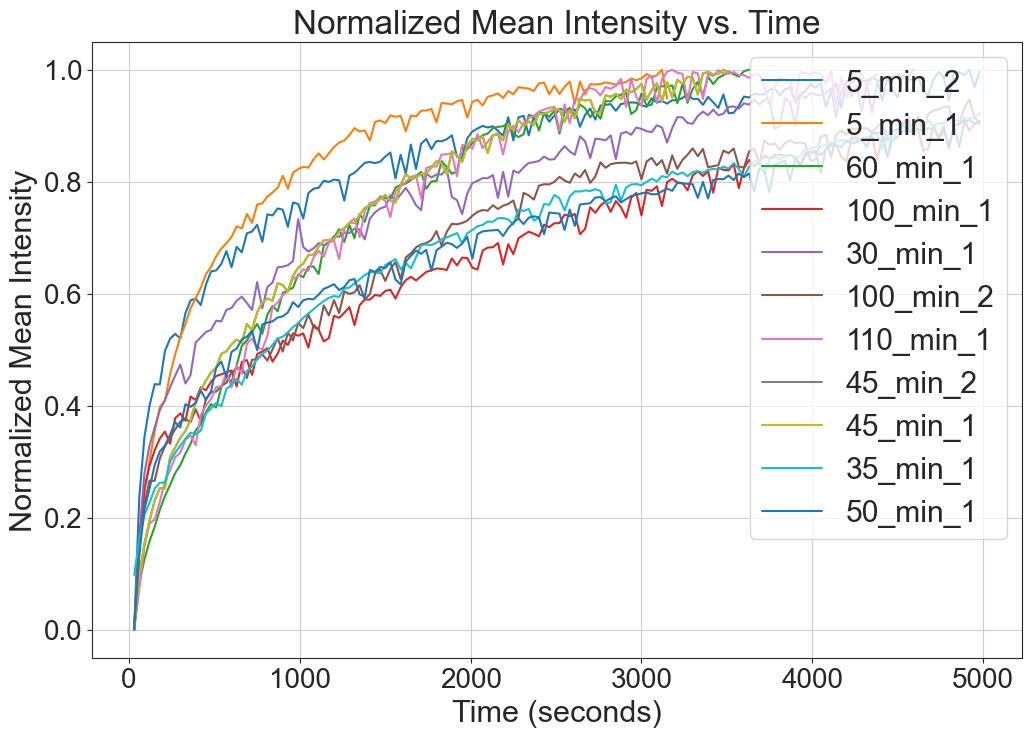}
\caption{
\textbf{Fluorescence Intensity During Recovery}}
\end{figure}
\subsection{Data Normalization}

Normalization of the fluorescence intensity values is performed to account for variability in initial fluorescence levels and photobleaching efficiency. The normalization is defined as:
\begin{equation}
I_{\text{normalized}}(t) = \frac{I(t) - I_{\text{min}}}{I_{\text{max}} - I_{\text{min}}},
\end{equation}
where $I(t)$ is the intensity at time $t$, $I_{\text{min}}$ is the minimum intensity, and $I_{\text{max}}$ is the maximum intensity observed in the recovery phase. This normalization ensures that the fluorescence intensity values range between 0 and 1.

\subsection{Half-time Recovery Calculation}

The half-time recovery ($\tau_{1/2}$) is a critical parameter in FRAP analysis, representing the time required for the fluorescence intensity to recover to half of its final value. Mathematically, it is determined as:
\begin{equation}
\tau_{1/2} = \min \left\{ t \mid I_{\text{normalized}}(t) \geq 0.5 \right\}.
\end{equation}

This value is identified by finding the time point at which the normalized intensity reaches 50

\subsection{Diffusion Coefficient Calculation}

The diffusion coefficient $D$ quantifies the rate at which molecules diffuse within the bleached area. It is derived from the half-time recovery using the following equation from Axelrod's paper:
\begin{equation}
D = \frac{w^2}{4 \tau_{1/2}} \gamma_D,
\end{equation}
where $w$ is the radius of the laser beam used for photobleaching, and $\gamma_D = 0.88$ is a geometric correction factor for circular beams. The radius $w$ is calculated as:
\begin{equation}
w = \text{resolution} \times \text{pixels},
\end{equation}
with the resolution of the optical setup being $0.294 \, \mu\text{m/pixel}$ and the number of pixels being 390, giving:
\begin{equation}
w = 0.294 \, \mu\text{m/pixel} \times 390 \, \text{pixels} = 114.66 \, \mu\text{m}.
\end{equation}

Thus, the diffusion coefficient is:
\begin{equation}
D = \frac{(114.66 \, \mu\text{m})^2}{4 \tau_{1/2}} \times 0.88.
\end{equation}

\section{Stokes Force Calculation}

To calculate the forces exerted by the fluid on tracer particles, we employed Stokes' law. This method involved tracking the motion of tracer beads within the fluid, determining their velocities, and then calculating the forces based on the fluid's viscosity and the particles' radii.

\subsection{Particle Tracking}
We utilized the same particle tracking code as described in the previous section, but with 10 µm tracer beads introduced into the system. The positions of these beads, $\mathbf{x}_i(t)$, were tracked over time $t$, and their velocities, $\mathbf{v}_i(t)$, were computed by differentiating their trajectories with respect to time:
\begin{equation}
\mathbf{v}_i(t) = \frac{d\mathbf{x}_i(t)}{dt}.
\end{equation}

\subsection{Velocity Calculation}
The velocities of the tracer beads were initially calculated in micrometers per second ($\mu$m/s). These velocities were then converted to meters per second (m/s) for use in the Stokes' law calculation:
\begin{equation}
v_{i, \text{m/s}} = v_{i, \mu\text{m/s}} \times 10^{-6}.
\end{equation}

\subsection{Stokes' Law}
Stokes' law relates the force $F$ exerted by a viscous fluid on a spherical particle moving through the fluid to the particle's radius $r$, the fluid's viscosity $\eta$, and the particle's velocity $v$. The equation is given by:
\begin{equation}
F = 6 \pi \eta r v.
\end{equation}

For our system, we used the following parameters:
\begin{itemize}
    \item Radius of the tracer particles: $r = 6.5 \times 10^{-6}$ meters.
    \item Viscosity of the fluid: $\eta = 2$ Pa.s.
\end{itemize}

The calculated forces for each particle were then averaged to obtain a representative value of the force in the system. The results, including velocities in both $\mu$m/s and m/s, and the corresponding forces in Newtons (N), were compiled into a DataFrame for further analysis and plotting.

\subsection{Example Calculation}
For clarity, an example calculation is provided below. Given a tracer bead velocity of $10 \, \mu\text{m/s}$, the force exerted on the particle is calculated as follows:
\begin{align*}
v_{i, \text{m/s}} &= 10 \times 10^{-6} \, \text{m/s}, \\
F_i &= 6 \pi \times 2 \times 6.5 \times 10^{-6} \times 10 \times 10^{-6}, \\
F_i &= 2.44 \times 10^{-9} \, \text{N}.
\end{align*}

\section{Percolation Theory Simulation}

We employ a percolation theory-based analogy to investigate the behavior of microtubule-kinesin active matter systems. The primary goal is to understand how the properties of the system, particularly the length of the microtubules, influence the formation of giant components, which can provide insights into the overall connectivity and functionality of the system. Code implementation in "percolation.ipynb" an simpler version could be found \hyperlink{https://colab.research.google.com/drive/1cwtlODipCBiXYNOKv2LF06Ri3byuxpaC?usp=sharing}{here}

\subsection{Mathematical Background}
Percolation theory is a mathematical framework that deals with the study of connected components in random graphs. In our analogy, we consider the microtubules as edges in a random graph, and the formation of giant components corresponds to the emergence of large-scale connectivity in the microtubule-kinesin system.

We model the lengths of the microtubules using a clamped gamma distribution, which is a continuous probability distribution defined by two parameters: shape ($k$) and scale ($\theta$), with an upper limit (clamp) at 25. The probability density function (PDF) of the clamped gamma distribution is given by:
\begin{equation}
f(x; k, \theta, \text{clamp}) = \frac{1}{\Gamma(k, \text{clamp}/\theta) \theta^k} x^{k-1} e^{-x/\theta}, \quad \text{for } 0 < x < \text{clamp}
\end{equation}
where $\Gamma(k, \text{clamp}/\theta)$ is the upper incomplete gamma function, defined as:
\begin{equation}
\Gamma(k, \text{clamp}/\theta) = \int_{0}^{\text{clamp}/\theta} x^{k-1} e^{-x} , dx
\end{equation}
The mean and variance of the clamped gamma distribution can be calculated numerically using the truncated moments.

\subsection{Simulation Methodology}
To investigate the formation of giant components in the microtubule-kinesin system, we perform the following steps:
\begin{enumerate}
    \item Generate microtubule lengths: We use the \texttt{generate\_microtubule\_lengths} function to generate a set of microtubule lengths based on the clamped gamma distribution with specified shape and scale parameters, and a clamp value of 25. The function ensures that the sum of the generated lengths is equal to a desired total length, which is set to approximately 1.
    \item Generate random graphs: We create random graphs using the \texttt{generate\_graph} function, where each node represents a microtubule and edges are added based on the edge probability distribution derived from the clamped gamma distribution. The edge probability is compared to a connection probability threshold ($p$) to determine whether an edge is added between two nodes.
    \item Analyze giant component ratio: The \texttt{analyze\_graph} function is used to calculate the giant component ratio, which is the ratio of the size of the largest connected component to the total number of nodes in the graph. This ratio serves as a measure of the overall connectivity of the system.
    \item Parameter exploration: We explore the parameter space by varying the shape and scale values of the clamped gamma distribution. For each combination of shape and scale, we generate multiple random graphs and calculate the average giant component ratio. This allows us to investigate how different microtubule length distributions affect the formation of giant components.
\end{enumerate}

\subsection{Visualization}
To visualize the results, we generate two main plots:
\begin{enumerate}
    \item Phase diagram: We create a heatmap-like plot where the x-axis represents the shape parameter and the y-axis represents the scale parameter of the clamped gamma distribution. The color of each point in the plot indicates the average giant component ratio for the corresponding shape and scale values. This plot provides an overview of the system's behavior across the parameter space.
    \item Representative graphs: We select a fixed shape value and vary the scale parameter to generate representative graphs of the microtubule-kinesin system. These graphs are visualized using a spring layout, where nodes represent microtubules and edges represent connections between them. The graphs provide a visual representation of the connectivity patterns that emerge under different conditions.
\end{enumerate}

By employing this percolation theory-based analogy and exploring the parameter space of the clamped gamma distribution, we gain valuable insights into the factors that influence the formation of giant components in microtubule-kinesin active matter systems. The mathematical framework and simulation methodology presented here offer a powerful tool for understanding the complex behavior of these systems and can guide future experimental and theoretical investigations.

\section{Cytosim Simulation}

\subsection{Mathematical Background and Cytosim Framework}
Cytosim is a powerful software package designed to simulate the collective dynamics of cytoskeletal fibers, such as microtubules and actin filaments \cite{Nedelec2007-nd}. It uses a coarse-grained representation of fibers and includes various biophysical processes, such as fiber flexibility, polymerization, depolymerization, and motor protein interactions. The mathematical framework underlying Cytosim is based on Langevin dynamics, which describes the motion of fibers and motors in a viscous medium subject to thermal fluctuations.
In Cytosim, fibers are modeled as inextensible elastic rods, which are discretized into segments. The dynamics of each segment $i$ is governed by the Langevin equation:
\begin{equation}
\gamma \frac{d\mathbf{r}_i}{dt} = \mathbf{F}_i^{\text{elastic}} + \mathbf{F}_i^{\text{motor}} + \mathbf{F}_i^{\text{thermal}} + \mathbf{F}_i^{\text{constraint}}
\end{equation}
where $\gamma$ is the drag coefficient, $\mathbf{r}_i$ is the position of segment $i$, and $\mathbf{F}_i$ represents the forces acting on the segment. The elastic forces $\mathbf{F}_i^{\text{elastic}}$ are derived from the bending and twisting energies of the fibers, while the motor forces $\mathbf{F}_i^{\text{motor}}$ are described by a stochastic binding and unbinding kinetics. The thermal forces $\mathbf{F}_i^{\text{thermal}}$ are modeled as Gaussian white noise, and the constraint forces $\mathbf{F}_i^{\text{constraint}}$ ensure the inextensibility of the fibers and the connectivity between segments.
The Langevin equation is solved numerically using a semi-implicit integration scheme:
\begin{equation}
\mathbf{r}_i(t + \Delta t) = \mathbf{r}_i(t) + \frac{\Delta t}{\gamma}\left(\mathbf{F}_i^{\text{elastic}} + \mathbf{F}_i^{\text{motor}} + \mathbf{F}_i^{\text{thermal}} + \mathbf{F}_i^{\text{constraint}}\right)
\end{equation}
where $\Delta t$ is the time step. This scheme ensures numerical stability and allows for the efficient simulation of large systems with many fibers and motors.
\subsection{Generating Cytosim Configuration Files}

To run simulations in Cytosim, users need to provide configuration files that specify the properties and initial conditions of the cytoskeletal components. In this study, we focus on the role of microtubule length distribution in the collective dynamics of microtubule-kinesin systems. We generate Cytosim configuration files with different microtubule length distributions based on the gamma distribution, which is characterized by two parameters: shape ($k$) and scale ($\theta$).

We define a function \texttt{generate\_microtubule\_lengths(shape, scale, total\_length)} that generates a set of microtubule lengths following a gamma distribution with the specified shape and scale parameters. The function also takes a \texttt{total\_length} argument, which determines the target sum of all generated microtubule lengths. The generated lengths are rounded to one decimal place and constrained to be between 1 and 25 units, which is a biologically relevant range for microtubule lengths.

The generated microtubule lengths are then used to create microtubule objects in the Cytosim configuration files. For each unique length value in the generated lengths list, a microtubule object is created with the corresponding length and the number of microtubules with that length. For example:

\begin{verbatim}
new 5 microtubule
{
length = 1.5
}
new 3 microtubule
{
length = 2.3
}
\end{verbatim}

In this example, there are 5 microtubules with a length of 1.5 and 3 microtubules with a length of 2.3.

To explore the effect of the gamma distribution parameters on the microtubule lengths and the resulting collective dynamics, we generate config files for two different scenarios: a 1D parameter sweep and a 2D parameter sweep. In the 1D sweep, we fix the shape parameter and vary the scale parameter, while in the 2D sweep, we vary both parameters simultaneously. Code implementation can be found in 1d-phase-diagram.ipynb and 2d-phase-diagram.ipynb, and manual-config.ipynb

\section{Microtubule Aster Tracking}

Tracking microtubule asters involves detecting and tracking their positions over time in image sequences. To improve detection accuracy, variable thresholding is applied, which adjusts the thresholding percentile based on the image characteristics, ensuring consistent detection across frames with varying intensity distributions. "figure\textunderscore 4\textunderscore aster \textunderscore traj.ipynb"

\subsection{Preprocessing with Variable Thresholding}

Variable thresholding adapts the threshold value dynamically by calculating the percentile of pixel intensities in each frame. This method helps in distinguishing microtubules from the background more effectively. The preprocessing steps are:

\begin{enumerate}
    \item Calculate the threshold value as a specified percentile of pixel intensities.
    \item Create a binary mask where pixels above the threshold are set to 255 (white) and others to 0 (black).
    \item Apply binary dilation to enhance the visibility of microtubule structures.
\end{enumerate}

For different segments of the image sequence, different percentiles are used:
\begin{itemize}
    \item The first 20 frames use a 99.4 percentile threshold.
    \item Frames 21 to 30 use a 99.5 percentile threshold.
    \item Frames 31 and beyond use a 99.97 percentile threshold.
\end{itemize}

\subsection{Aster Tracking}

After preprocessing, the TrackPy library is used to detect and track microtubule asters. The tracking steps are:

\begin{enumerate}
    \item Detect particles in each frame using a bandpass filter.
    \item Link particles across frames to form trajectories based on proximity and predicted motion.
    \item Calculate velocities by differentiating trajectories with respect to time.
\end{enumerate}

This method allows for accurate tracking of microtubule asters, providing valuable data on their dynamics under different experimental conditions.

\subsection{Aster Area Calculating}

Aster area can be computed by calculating the area above the intensity threshold. Implemented in "figure\textunderscore 4\textunderscore aster\textunderscore size.ipynb"

\section{Cell Tracking and Velocity Calculation}

We tracked cells in brightfield images to analyze their movements and calculate associated physical quantities such as velocity, force, work, and power. The tracking was performed using a similar approach to particle tracking. Code implementation in "figure\textunderscore 4\textunderscore taj.ipynb"

\subsection{Cell Detection and Tracking}

Cells were imaged using brightfield microscopy. The detection and tracking process involved the following steps:

\begin{enumerate}
    \item \textbf{Detection:} Cells were identified in each frame based on intensity and morphological criteria.
    \item \textbf{Linking:} Detected cells were linked across frames to form trajectories, ensuring that each cell's movement was accurately captured over time.
\end{enumerate}

The position $\mathbf{x}_i(t)$ of cell $i$ at time $t$ was recorded, allowing for the calculation of displacement and velocity.

\subsection{Velocity Calculation}

The velocity $\mathbf{v}_i(t)$ of cell $i$ was calculated by differentiating its trajectory with respect to time.

\begin{equation}
\text{velocity} = \text{velocity} \times 0.43 \, \text{µm/pixel}
\end{equation}

\subsection{Tracer Bead Force Calculation}

The force exerted by or on the cells was calculated using Stokes' law, previously described in section 5. In short, we calculate the force as the force experienced by a spherical particle moving through a viscous fluid to its velocity:

\begin{equation}
F = 6 \pi \eta r v,
\end{equation}
where:
\begin{itemize}
    \item $F$ is the force,
    \item $\eta$ is the viscosity of the fluid,
    \item $r$ is the radius of the cell,
    \item $v$ is the velocity of the cell.
\end{itemize}

In this case, we used a radius $r = 6.5 \times 10^{-6}$ meters (6.5 µm) and fluid viscosity $\eta = 2 \times 10^{-3}$ Pa.s. The velocity $v$ was converted from µm/s to m/s using the pixel-to-meter conversion factor.

\subsection{Work and Power Calculation}

The work done by the cells was calculated by integrating the force over the displacement for each time step:

\begin{equation}
\text{work\_done} = F \cdot \text{displacement},
\end{equation}
where the displacement was converted from pixels to meters.

The total work done over the entire observation period was obtained by summing the work done at each time step.

\begin{equation}
\text{total\_work\_done} = \sum \text{work\_done}
\end{equation}

The power, which is the rate of doing work, was calculated by dividing the work done at each time step by the corresponding time interval:

\begin{equation}
\text{power} = \frac{\text{work\_done}}{\Delta t},
\end{equation}
where $\Delta t$ is the time between frames (1/8 seconds).

The total power at each time step was then summed to obtain the overall power profile as a function of time, which was plotted to visualize the energy expenditure of the cells.

\begin{equation}
\text{total\_power} = \sum \text{power}
\end{equation}

The power was plotted against time to show how the energy expenditure varied over the course of the experiment.

\section{Experimental Setup}
\subsection{Light-induced kinesin expression and purification}
We constructed two chimeras of D. melanogaster kinesin K401: K401-iLID and K401-micro as previously described (Addgene 122484 and 122485) \cite{Ross2019-ol}. In short, for the K401-iLiD plasmid we inserted iLID with a His tag after the C-terminus of K401. For the K401-micro plasmid, we inserted K401 between the His-MBP and the micro. The MBP domain is needed to ensure the microdomain remains fully functional during expression \cite{Guntas2015-to}. After the expression, the MBP domain can be cleaved of by TEV protease.

For protein expression, we transformed the plasmids to BL21(DE3)pLysS cells. The cells were grown in LB and induced at with 1mM IPTG at 18°C for 16 hours after reaching OD 0.6. The cells were then pelleted at 4000G and resuspended in lysis buffer (50 mM sodium phosphate, 4 mM MgCl2, 250 mM NaCl, 25 mM imidazole, 0.05mM MgATP, 5 mM BME, 1 mg/ml lysozyme and 1 tablet/50 mL of Complete Protease Inhibitor). After a 1-hour incubation with stirring, the lysate was passed through a 30kPSI cell disruptor. The lysate was clarified at 30,000 G for 1 hour. The supernatant was then incubated with Ni-NTA agarose resin for 1 hour. The lysate/Ni-NTA mixture was loaded into a chromatography column and washed three times with wash buffer (Lysis buffer with no lysozyme nor complete EDTA tablet), and eluted with 500mM imidazole. Protein elutions were dialyzed overnight using 30 kDa MWCO membrane against 50 mM sodium phosphase, 4 mM MgCl2, 250 mM NaCl, 0.05 mM MgATP, and 1 mM BME. For the K401-micro elution, we added TEV protease at a 1:25 mass ratio to remove the MBP domain. Then, we used centrifugal filters to exchange to pH 6.7 protein storage buffer (50 mM imidazole, HCl for pH balancing, 4 mM MgCl2, 2 mM DTT, 50 µM MgATP, and 36\% sucrose). Proteins were then aliquoted and flash frozen in LN2 and stored under -80°C.

\subsection{Microtubule Polymerization and Length Distribution}

Fluorescent microtubule polymerization was previously described \cite{Ross2019-ol}. In short, we used a protocol based on one found on the Dogic lab homepage. The procedure began by preloading and starting a 37℃ water bath. GMP-cpp, reagents, and tubes were cooled on ice. A 20 mM DTT solution was prepared using Pierce no-weigh format, and a GMP mixture consisting of M2B, DTT, and GMP-cpp was made and stored on ice. The ultracentrifuge and rotor were pre-cooled to 4℃. Tubulin (20 mg/mL) and labeled tubulin (20 mg/mL) were thawed in the water bath until mostly thawed and then cooled on ice. In a cold room, labeled tubulin was added to the stock vial of unlabeled tubulin and mixed gently. The GMP mixture was then added to the tubulin mixture and stirred gently. This combined mixture was pipetted into ultracentrifuge tubes and incubated on ice for 5 minutes before being centrifuged at 90,000 rpm, 4℃ for 8 minutes. The supernatant was carefully collected without disturbing the pellet and transferred to an Eppendorf tube, mixed, and stored on ice. The mixture was incubated in a 37℃ water bath for 1 hour, protected from light. Aliquots were then dispensed into PCR strip tubes, which were spun to collect the fluid at the bottom. Finally, the PCR strips were flash frozen in liquid nitrogen and stored in a -80℃ freezer.

To measure the length distribution of microtubules, we imaged fluorescently labeled microtubules immobilized onto the cover glass surface of a flow cell. The cover glass was treated with a 0.01\% solution of poly-L-lysine (Sigma P4707) to promote microtubule binding. The lengths of microtubules were determined by image segmentation. Each microtubule image was normalized and underwent local and global thresholding to correct for non-uniform backgrounds and obtain thresholded images of putative microtubules. Morphological operations were applied to reconnect small breaks in filaments. Objects near the image boundary were removed, and small or circular objects were filtered out based on size and eccentricity thresholds. Potential microtubule crossovers were identified and eliminated by analyzing the angles of lines within the image.

\subsection{Sample Chambers for Experiments}
Glass slides and coverslips were first cleaned using a series of washes. Slides and coverslips were placed in respective containers, and 2\% Hellmanex solution was prepared by mixing 6 mL Hellmanex with 300 mL DI water, heated, and poured into the containers. The containers were sonicated for 10 minutes, followed by three DI water rinses and an ethanol rinse. Ethanol was added to the containers and sonicated again for 10 minutes, followed by another ethanol rinse and three DI water rinses. Next, 0.1 M KOH was added to the containers, sonicated for 10 minutes, and rinsed three times with DI water. The slides were then etched overnight with 5\% HCl and rinsed three times with DI water. Clean slides and coverslips were stored in DI water.

For silane coupling, a 2\% acrylamide solution was prepared using 40\% acrylamide stock solution, and degassed under vacuum. In a chemical hood, 98.5\% ethanol, 1\% acetic acid, and 0.5\% silane agent were mixed to prepare the silane-coupling solution, which was immediately poured into the containers with the slides and coverslips and incubated at room temperature for 20-30 minutes. The slides were then rinsed once with ethanol, three times with DI water, and baked at 110°C for 30 minutes or 50°C overnight.

For acrylamide polymerization, the degassed 300 mL of 2\% acrylamide solution was moved to a stir plate, and 105 µL TEMED and 210 mg ammonium persulfate were added. The solution was immediately poured over the silane-coupled slides and coverslips and left to polymerize overnight at 4°C. Before use, the slides and coverslips were rinsed with DI water and air-dried.

\subsection{Reaction Mixture}
For the self-organization experiments, K401-micro, K401-iLID, and microtubules were combined into a reaction mixture to achieve final concentrations of approximately 0.1 µM for each motor type and 1.5-2.5 µM for tubulin, referring to protein monomers for K401-micro and K401-iLID constructs, and protein dimers for tubulin. The sample preparation was conducted under dark-room conditions to minimize unintended light activation, using room light filtered to block wavelengths below 580 nm (Kodak Wratten Filter No. 25). The base reaction mixture included a buffer, MgATP as an energy source, glycerol as a crowding agent, pluronic F-127 for surface passivation, and components for oxygen scavenging (pyranose oxidase, glucose, catalase, Trolox, DTT), along with ATP-recycling reagents (pyruvate kinase/lactic dehydrogenase, phosphoenolpyruvic acid). The reaction mixture consisted of 59.2 mM K-PIPES pH 6.8, 4.7 mM MgCl2, 3.2 mM potassium chloride, 2.6 mM potassium phosphate, 0.74 mM EGTA, 1.4 mM MgATP, 10\% glycerol, 0.50 mg/mL pluronic F-127, 2.9mg/mL pyranose oxidase, 3.2 mg/mL glucose, 0.086 mg/mL catalase, 5.4 mM DTT, 2.0 mM Trolox, 0.026 units/µL pyruvate kinase/lactic dehydrogenase, and 26.6 mM phosphoenolpyruvic acid.

We note that the sample is sensitive to the buffer pH and mixture incubation time. For our experimental conditions, the mixture pH is around 6.4 and we perform the experiments within 2 hours of constructing the mixture.

\subsection{Microscope Setup}

We conducted the experiments using an automated widefield epifluorescence microscope (Nikon Ti-2), custom-modified for two additional imaging modes: epi-illuminated pattern projection and LED-gated transmitted light. Light patterns from a programmable DLP chip (EKB Technologies DLP LightCrafter™ E4500 MKII™ Fiber Couple) were projected onto the sample via a user-modified epi-illumination attachment (Nikon T-FL). The DLP chip was illuminated by a fiber-coupled 470 nm LED (ThorLabs M470L3). The epi-illumination attachment featured two light-path entry ports: one for the projected pattern light path and the other for a standard widefield epi-fluorescence light path. These light paths were combined using a dichroic mirror (Semrock BLP01-488R-25). The magnification of the epi-illumination system was calibrated to ensure the camera's imaging sensor (FliR BFLY-U3-23S6M-C) was fully illuminated when the entire DLP chip was activated. Micro-Manager software, running custom scripts, controlled the pattern projection and stage movement. For the transmitted light path, we replaced the standard white-light brightfield source (Nikon T-DH) with an electronically time-gated 660 nm LED (ThorLabs M660L4-C5) to minimize light-induced dimerization during brightfield imaging.

\subsection{Tracer Particle Method for Measuring Fluid Velocity}

To measure the fluid velocity, we employed 1 µm polystyrene beads (Polysciences 07310-15) as tracer particles. The hydrophobic surface of these beads was passivated by incubating them overnight in M2B buffer with 50 mg/ml of pluronic F-127. This incubation process helps to coat the beads with pluronic, which prevents nonspecific interactions with the surrounding fluid and ensures accurate velocity measurements.

Before each experiment, the pluronic-coated beads were thoroughly washed. This was achieved by pelleting the beads through centrifugation and subsequently resuspending them in M2B buffer containing 0.5 mg/ml pluronic. This concentration of pluronic was chosen to match that of the reaction mixture, thereby maintaining consistency in the experimental conditions.

The beads, once prepared, were introduced into the fluid of interest. Their motion was tracked using high-resolution imaging techniques, and their velocities were calculated by analyzing their trajectories over time. This method provides precise measurements of the fluid velocity by leveraging the well-characterized motion of the tracer particles.

\subsection{Culturing Jurkat Cells Using RPMI Medium}

Jurkat cells, a type of immortalized T lymphocyte cell line, were cultured using RPMI 1640 medium. The RPMI 1640 medium provides the necessary nutrients and environment for the optimal growth and maintenance of Jurkat cells.

The cells were maintained in RPMI 1640 medium supplemented with the following components to enhance cell viability and proliferation:
\begin{itemize}
    \item 10\% fetal bovine serum (FBS) to provide essential growth factors, hormones, and proteins.
    \item 1\% penicillin-streptomycin to prevent bacterial contamination.
    \item 2 mM L-glutamine to support protein and nucleotide synthesis.
\end{itemize}

The culturing process involved the following steps:
\begin{enumerate}
    \item \textbf{Cell Seeding:} Jurkat cells were seeded into culture flasks at an initial density of $0.2 \times 10^6$ cells/ml.
    \item \textbf{Incubation:} The flasks were incubated at 37°C in a humidified atmosphere containing 5\% CO\textsubscript{2}. This environment mimics the physiological conditions of the human body, promoting optimal cell growth.
    \item \textbf{Medium Change:} The culture medium was replaced every 2-3 days to remove metabolic waste products and replenish nutrients. This was done by gently pelleting the cells through centrifugation, discarding the old medium, and resuspending the cells in fresh RPMI 1640 medium with supplements.
    \item \textbf{Cell Passage:} When the cells reached a density of $1 \times 10^6$ cells/ml, they were subcultured to maintain optimal growth conditions. This involved diluting the cell suspension with fresh medium and seeding it into new culture flasks.
\end{enumerate}

We also tested cell viability after prolonged incubation in active matter mix using Trypan Blue. Jurkat cells showed minimal cell death during the first 180 minutes of incubation in active matter mixture. 

\subsection{Active Matter in Oil Emulsion}

To prepare the active matter in oil emulsion using the EURx MICELLULA DNA Emulsion kit, follow these steps:

\begin{enumerate}
    \item \textbf{Prepare the Oil Emulsion Mix}
    \begin{tabular}{|c|c|}
        \hline
        \textbf{Component} & \textbf{Volume (µL)} \\
        \hline
        Component 1 & 22 \\
        \hline
        Component 2 & 2 \\
        \hline
        Component 3 & 6 \\
        \hline
    \end{tabular}

    \item \textbf{Combine the Reaction Mix with the Oil Emulsion Mix}
    \begin{itemize}
        \item Add 5 µL of the reaction mix to 30 µL of the prepared oil emulsion mix.
    \end{itemize}

    \item \textbf{Create a Stable Emulsion}
    \begin{itemize}
        \item Run the tube containing the mixture against a tube rack to create a stable emulsion.
        \item Vortex the mixture if necessary to ensure proper emulsification.
    \end{itemize}
\end{enumerate}

\section{Code Avilibility}
Code can be found at \hyperlink{https://github.com/shichenliu97/active-matters-programmed-phase-transition}{https://github.com/shichenliu97/active-matters-programmed-phase-transition}

\bibliographystyle{unsrt}
\bibliography{selforg}